\documentclass[a4paper,UKenglish]{lipics-v2016}

\pdfoutput=1

\usepackage{amsthm}

\usepackage{amssymb}
\usepackage{amsmath}
\usepackage{times}

\theoremstyle{plain}
\newtheorem{The} {Theorem} [section]
\newtheorem{Cor} [The] {Corollary}
\newtheorem{lem} [The] {Lemma}

\newtheorem{Rem}[The]{Remark}
\newtheorem{Exa}[The]{Example}

\newtheorem{Deff}[The]{Definition}

\newcommand{\borapxi}{{\bf\Sigma}^{0}_{\xi}}

\newcommand{\bormpxi}{{\bf\Pi}^{0}_{\xi}}

\newcommand{\borel}{{\bf\Delta}^{1}_{1}}

\newcommand{\hs}{\hspace{12mm}

}
\newcommand{\noi}{\noindent}

\newcommand{\om}{\omega}
\newcommand{\Si}{\Sigma}
\newcommand{\Sis}{\Sigma^\star}
\newcommand{\Sio}{\Sigma^\omega}
\newcommand{\nl}{\newline}
\newcommand{\lra}{\leftrightarrow}
\newcommand{\fa}{\forall}
\newcommand{\ra}{\rightarrow}

\newcommand{\Exab}{\begin{Exa}}
\newcommand{\Exae}{\end{Exa}}

\newcommand{\Ga}{\Gamma}
\newcommand{\Gas}{\Gamma^\star}
\newcommand{\Gao}{\Gamma^\omega}
\newcommand{\ite}{\item}

\newcommand{\ol}{$\omega$-language}

\newcommand{\proo}{\noi {\bf Proof.} }
\newcommand {\ep}{\hfill $\square$}

\title{Wadge Degrees  of $\om$-Languages of Petri Nets}
\titlerunning{Wadge Degrees  of $\om$-Languages of Petri Nets}

\author[]{Olivier Finkel}

\affil[]{ Equipe de Logique Math\'ematique
\\Institut de Math\'ematiques de Jussieu - Paris Rive Gauche
 \\  CNRS et Universit\'e Paris 7, France. \\ 
  \texttt{finkel@math.univ-paris-diderot.fr}}

\authorrunning{O. Finkel}

\Copyright{Olivier Finkel}


\begin{document}

\maketitle

\begin{abstract}
We prove that  $\om$-languages of (non-deterministic) Petri nets and $\om$-languages of (non-deterministic)  Turing machines   have the same topological complexity:  
 the Borel and Wadge hierarchies of the class of  $\om$-languages of (non-deterministic) Petri nets are equal to the Borel and Wadge 
hierarchies of the class of $\om$-languages of (non-deterministic) Turing machines which also form the class of effective analytic sets. In particular,  for each non-null recursive ordinal 
$\alpha < \om_1^{{\rm CK}} $ there 
exist some ${\bf \Si}^0_\alpha$-complete and some 
 ${\bf \Pi}^0_\alpha$-complete   $\om$-languages of Petri nets, 
and the supremum 
of the set of Borel ranks of $\om$-languages  of Petri nets is the ordinal $\gamma_2^1$, 
which  is strictly greater than the first non-recursive ordinal $\om_1^{{\rm CK}}$. 
We also prove that there are some ${\bf \Si}_1^1$-complete, hence non-Borel, $\om$-languages of Petri nets, and that it is consistent with ZFC that there exist some 
$\om$-languages of Petri nets which are neither Borel nor  ${\bf \Si}_1^1$-complete. 
This  answers  the question of the topological complexity of $\om$-languages of (non-deterministic) Petri nets which was left open in \cite{DFR4,FS14}.

\end{abstract}

\section{Introduction}

In the sixties, B\"uchi was the first to study acceptance of  infinite words by finite automata with the now called  B\"uchi  acceptance condition, in order 
to prove the decidability of the monadic second order theory of one successor
over the integers.  Since then there has been a lot of work on  regular $\omega$-languages, accepted by   B\"uchi automata, or by some other variants of automata over infinite words, like Muller or Rabin automata,
see \cite{Thomas90,Staiger97,PerrinPin}. 
The acceptance of infinite words by other   finite machines, like 
pushdown automata, counter automata, Petri nets, Turing machines, \ldots, with various acceptance conditions, 
 has also been studied,  see  \cite{Staiger97,eh,CG78b,valk1983infinite,Staiger86a,Staiger87b,Staiger93}. 
 
The Cantor topology is a very natural topology on the set $\Si^\omega$ of infinite words over a finite alphabet $\Si$ which is induced by the prefix metric.  Then a
 way to study the complexity of   languages of infinite words  
accepted by finite machines          is to study    their      topological complexity and firstly  
to locate them with regard to 
the Borel and the projective hierarchies \cite{Thomas90,eh,LescowThomas,Staiger97}.  

Every $\omega$-language accepted by a deterministic B\"uchi automaton is a ${\bf \Pi}^0_2$-set. 
On the other hand,   it follows from Mac Naughton's Theorem that every regular 
 $\omega$-language    is  accepted by a   deterministic
Muller   automaton, and thus is a boolean combination of $\omega$-languages accepted by  deterministic B\"uchi automata. Therefore 
every regular $\omega$-language  is a 
${\bf \Delta}^0_3$-set. Moreover Landweber proved that  
the Borel complexity of any $\omega$-language accepted by a  Muller or B\"uchi  automaton can be effectively  computed 
(see \cite{Landweber69,PerrinPin}).
In a similar way,  every $\omega$-language accepted by a deterministic Muller Turing machine, and thus also by any 
Muller deterministic finite machine is a ${\bf \Delta}^0_3$-set, \cite{eh,Staiger97}. 

On the other hand, the Wadge hierarchy is a great refinement of the Borel hierarchy, firstly defined by Wadge via reductions by continuous functions \cite{Wadge83}. 
 The trace of the Wadge hierarchy on the $\omega$-regular languages is called the Wagner hierarchy. It has been completely described by 
Klaus Wagner in \cite{Wagner79}. Its length is the ordinal $\omega^\omega$.
 Wagner gave an automaton-like characterization of this hierarchy, based on the notions of chain
 and superchain, together with an 
algorithm to compute the Wadge (Wagner) degree of any given  $\omega$-regular language, see also 
  \cite{CartonPerrin97b,CartonPerrin99,Selivanov98,Selivanov08m,Selivanov08t,Simonnet92}.

The Wadge  hierarchy of  deterministic context-free  $\omega$-languages was  determined by Duparc in \cite{Duparc03,dfr}. Its length is the ordinal 
  $\omega^{(\omega^2)}$. 
We do not know yet whether this hierarchy is decidable or not. 
 But the Wadge hierarchy induced by deterministic partially blind 
  1-counter automata was described in an effective way in \cite{Fin01csl}, and other partial decidability results were obtained in \cite{Fin01a}. 
Then, it was proved in  \cite{Fin-mscs06} that the Wadge  hierarchy of  $1$-counter or context-free 
 $\omega$-languages and the Wadge hierarchy of effective analytic sets (which form the class of all the $\omega$-languages accepted by non-deterministic Turing machines)  
 are equal. 
Moreover   similar results hold about the Wadge hierarchy of infinitary rational relations  accepted by $2$-tape B\"uchi automata, \cite{Fink-Wd}.  
 Finally, the Wadge hierarchy of $\omega$-languages of deterministic 
Turing machines was determined by Selivanov in \cite{Selivanov03b}.

We consider in this paper acceptance of infinite words by Petri nets. 
Petri nets are used for the description of distributed systems  \cite{esparza1998decidability,rozenberg2004lectures,H-petrinet-diaz}, and form a  very important  mathematical model in Concurrency Theory 
that has been developed for general concurrent computation. 
In the context of Automata Theory,  Petri nets may be defined as 
(partially) blind multicounter automata, as explained in \cite{valk1983infinite,eh,Gre78}. 
First,  one can distinguish between the 
places of a given Petri net by dividing them  into the bounded ones (the number of tokens 
in such a place at any time is uniformly bounded) and the unbounded ones. Then 
each unbounded place may be seen as a partially blind counter, and the tokens in the bounded places 
determine the state of the partially blind multicounter automaton that is equivalent to the initial Petri net. The transitions of the Petri net may then be seen as the finite control of the 
partially blind multicounter automaton and the labels of these transitions are then the input symbols. 
The infinite behavior of Petri nets was first studied by 
Valk \cite{valk1983infinite} and  by Carstensen in the case of deterministic Petri nets  \cite{carstensen1988infinite}.

On one side, the topological complexity of  $\om$-languages of  {\it deterministic}  Petri nets is completely determined. They  are  ${\bf \Delta}^0_3$-sets 
and their Wadge hierarchy
 has been determined by Duparc, Finkel and Ressayre in \cite{DFR4}; its length is the ordinal $\om^{\om^2}$.  
On the other side,  Finkel and Skrzypczak proved in \cite{FS14} that there exist  ${\bf \Sigma}_3^0$-complete, hence non  ${\bf \Delta}_3^0$,   $\om$-languages accepted by  {\it non-deterministic} one-partially-blind-counter B\"uchi automata. But, 
up to our knowledge, this was the only known result  about the  topological complexity of     $\om$-languages of  {\it non-deterministic}  Petri nets.   
Notice  that $\om$-languages accepted by (non-blind) one-counter 
B\"uchi  automata have the same topological complexity as 
$\om$-languages of Turing machines, \cite{Fin-mscs06}, but the non-blindness of the counter, i.e. the ability to use the zero-test of the counter,  was essential 
in the proof of this result.

Using a simulation of a  given real time 
$1$-counter (with zero-test) B\"uchi automaton $\mathcal{A}$ accepting 
$\om$-words $x$ over the alphabet $\Si$ by a real time 4-blind-counter B\"uchi automaton $\mathcal{B}$ reading some special codes $h(x)$ of  the words $x$,
we prove here    that  $\om$-languages of {\it non-deterministic}  Petri nets and effective analytic sets have the same topological complexity:  
 the Borel and Wadge hierarchies of the class of  $\om$-languages of  Petri nets are equal to the Borel and Wadge 
hierarchies of the class of effective analytic sets. In particular,  for each non-null recursive ordinal 
$\alpha < \om_1^{{\rm CK}} $ there 
exist some ${\bf \Si}^0_\alpha$-complete and some 
 ${\bf \Pi}^0_\alpha$-complete   $\om$-languages of Petri nets, 
and the supremum 
of the set of Borel ranks of $\om$-languages  of Petri nets is the ordinal $\gamma_2^1$, 
which  is strictly greater than the first non-recursive ordinal $\om_1^{{\rm CK}}$. 
We also prove that there are some ${\bf \Si}_1^1$-complete, hence non-Borel, $\om$-languages of Petri nets, and  that it is consistent with ZFC that there exist some 
$\om$-languages of Petri nets which are neither Borel nor  ${\bf \Si}_1^1$-complete. 

  The paper is organized as follows.  
 In Section 2 we review the  notions of (blind) counter automata and $\om$-languages. In Section 3 we recall notions of topology, and in particular the Borel and Wadge hierarchies on 
a Cantor space. We prove our main results 
  in Section 4. Concluding remarks are given in Section 5.

\section{Counter Automata}
 
 We assume   the reader to be familiar with the theory of formal ($\om$-)languages  
\cite{Staiger97,PerrinPin}.
We recall the  usual notations of formal language theory. 

If  $\Si$ is a finite alphabet, a {\it non-empty finite word} over $\Si$ is any 
sequence $x=a_1\ldots a_k$, where $a_i\in\Sigma$ 
for $i=1,\ldots ,k$, and  $k$ is an integer $\geq 1$. The {\it length}
 of $x$ is $k$, denoted by $|x|$.
 The {\it empty word}  is denoted by $\lambda$; its length is $0$. 
 $\Sis$  is the {\it set of finite words} (including the empty word) over $\Sigma$, and we denote $\Si^+=\Sis \setminus \{\lambda\}$.
A  (finitary) {\it language} $V$ over an alphabet $\Sigma$ is a subset of  $\Sis$.

 The {\it first infinite ordinal} is $\om$.
 An $\om$-{\it word} over $\Si$ is an $\om$-sequence $a_1 \ldots a_n \ldots$, where for all 
integers $ i\geq 1$, ~
$a_i \in\Sigma$.  When $\sigma=a_1 \ldots a_n \ldots$ is an $\om$-word over $\Si$, we write
 $\sigma(n)=a_n$,   $\sigma[n]=\sigma(1)\sigma(2)\ldots \sigma(n)$  for all $n\geq 1$ and $\sigma[0]=\lambda$.

 The usual concatenation product of two finite words $u$ and $v$ is 
denoted $u\cdot v$ (and sometimes just $uv$). This product is extended to the product of a 
finite word $u$ and an $\om$-word $v$: the infinite word $u\cdot v$ is then the $\om$-word such that:

 $(u\cdot v)(k)=u(k)$  if $k\leq |u|$ , and 
 $(u\cdot v)(k)=v(k-|u|)$  if $k>|u|$.
  
 The {\it set of } $\om$-{\it words} over  the alphabet $\Si$ is denoted by $\Si^\om$.
An  $\om$-{\it language} $V$ over an alphabet $\Sigma$ is a subset of  $\Si^\om$, and its  complement (in $\Sio$) 
 is $\Sio \setminus V$, denoted $V^-$.

  The {\it prefix relation} is denoted $\sqsubseteq$: a finite word $u$ is a {\it prefix} 
of a finite word $v$ (respectively,  an infinite word $v$), denoted $u\sqsubseteq v$,  
 if and only if there exists a finite word $w$ 
(respectively,  an infinite word $w$), such that $v=u\cdot w$.  

  Let $k$ be an integer $\geq 1$. 
A  $k$-counter machine has $k$ {\it counters}, each of which containing a  non-negative integer. 
The machine can test whether the content of a given counter is zero or not, but this is not possible if the counter is a blind (sometimes called partially blind, as in \cite{Gre78}) counter.  
This means that  if a transition of the machine is enabled  when the content of a blind counter 
is zero then the same transition is also enabled when the content of the same counter 
is a non-zero integer.
And transitions depend on the letter read by the machine, the current state of the finite control, and the tests about the values of the counters. 
Notice that in the sequel we shall only consider real-time automata, i.e. $\lambda$-transitions are not allowed (but the general results of this paper will be easily extrended to the case of non-real-time automata).

 Formally a   {\it real time} $k$-counter machine is a 4-tuple 
$\mathcal{M}$=$(K,\Si,$ $ \Delta, q_0)$,  where $K$ 
is a finite set of states, $\Sigma$ is a finite input alphabet, 
 $q_0\in K$ is the initial state, 
and  $\Delta \subseteq K \times  \Si\times \{0, 1\}^k \times K \times \{0, 1, -1\}^k$ is the transition relation.

If  the machine $\mathcal{M}$ is in state $q$ and 
$c_i \in \mathbf{N}$ is the content of the $i^{th}$ counter 
 $\mathcal{C}$$_i$ then 
the  configuration (or global state)
 of $\mathcal{M}$ is the  $(k+1)$-tuple $(q, c_1, \ldots , c_k)$.

 For $a\in \Si$, 
$q, q' \in K$ and $(c_1, \ldots , c_k) \in \mathbf{N}^k$ such 
that $c_j=0$ for $j\in E \subseteq  \{1, \ldots , k\}$ and $c_j >0$ for 
$j\notin E$, if 
$(q, a, i_1, \ldots , i_k, q', j_1, \ldots , j_k) \in \Delta$ where $i_j=0$ for $j\in E$ 
and $i_j=1$ for $j\notin E$, then we write:

~~~~~~~~$a: (q, c_1, \ldots , c_k)\mapsto_{\mathcal{M}} (q', c_1+j_1, \ldots , c_k+j_k)$.

 Thus  the transition relation must obviously satisfy:
 \nl if $(q, a, i_1, \ldots , i_k, q', j_1, \ldots , j_k)  \in    \Delta$ and  $i_m=0$ for 
 some $m\in \{1, \ldots , k\}$  then $j_m=0$ or $j_m=1$ (but $j_m$ may not be equal to $-1$). 
  
  Moreover if the  
counters of $\mathcal{M}$ are  blind, then, if $(q, a, i_1, \ldots , i_k, q', j_1, \ldots , j_k)  \in    \Delta$  holds, and  $i_m=0$ for 
 some $m\in \{1, \ldots , k\}$ then  $(q, a, i_1, \ldots , i_k, q', j_1, \ldots , j_k)  \in    \Delta$
    also holds if   $i_m=1$ and the other intergers are unchanged.

An $\om$-sequence of configurations $r=(q_i, c_1^{i}, \ldots c_k^{i})_{i \geq 1}$ is called 
a run of $\mathcal{M}$ on an $\om$-word  $\sigma=a_1a_2 \ldots a_n \ldots $   over $\Si$ iff:

(1)  $(q_1, c_1^{1}, \ldots c_k^{1})=(q_0, 0, \ldots, 0)$

(2)   for each $i\geq 1$, 

~~ $a_i: (q_i, c_1^{i}, \ldots c_k^{i})\mapsto_{\mathcal{M}}  
(q_{i+1},  c_1^{i+1}, \ldots c_k^{i+1})$.

For every such run $r$, $\mathrm{In}(r)$ is the set of all states entered infinitely
 often during $r$.

\begin{Deff} A B\"uchi $k$-counter automaton  is a 5-tuple 
$\mathcal{M}$=$(K,\Si,$ $\Delta, q_0, F)$, 
where $ \mathcal{M}'$=$(K,$ $\Si,$  $\Delta, q_0)$
is a $k$-counter machine and $F \subseteq K$ 
is the set of accepting  states.
The \ol~ accepted by $\mathcal{M}$ is:
$L(\mathcal{M})$= $\{  \sigma\in\Si^\om \mid \mbox{  there exists a  run r
 of } \mathcal{M} \mbox{ on } \sigma \mbox{  such that } \mathrm{In}(r)
 \cap F \neq \emptyset \}$

\end{Deff}

\begin{Deff} A Muller $k$-counter automaton  is a 5-tuple 
$\mathcal{M}$=$(K,$ $\Si, \Delta, q_0, \mathcal{F})$, 
 where $ \mathcal{M}'$=$(K,$ $\Si, \Delta, q_0)$
is a $k$-counter machine and $\mathcal{F}$$ \subseteq 2^K$ 
is the set of accepting sets of  states.
The \ol~ 
accepted by $\mathcal{M}$ is: 
$L(\mathcal{M})$=$ \{  \sigma\in\Si^\om \mid \mbox{  there exists a  run r
 of } \mathcal{M}$$ \mbox{ on } \sigma \mbox{  such that } 
\exists F \in \mathcal{F}$$ ~~ \mathrm{In}(r)=F  \}$

\end{Deff}

 It is well known that an $\om$-language is accepted by a non-deterministic (real time) 
B\"uchi $k$-counter automaton iff it is accepted by a non-deterministic
(real time) Muller  $k$-counter automaton \cite{eh}. Notice that it cannot be 
 shown without  using the 
non determinism of automata and this result is no longer true in the  deterministic case.

  The class of \ol s accepted by real time $k$-counter B\"uchi  automata  (respectively,   real time $k$-blind-counter B\"uchi  automata)      is  
denoted {\bf r}-${\bf CL}(k)_\om$. (respectively, {\bf r}-${\bf BCL}(k)_\om$). (Notice that in previous papers, as in \cite{Fin-mscs06}, the class {\bf r}-${\bf CL}(k)_\om$ was denoted  {\bf r}-${\bf BCL}(k)_\om$
so we have slightly changed the notation in order to distinguish the different classes). 

  The class ${\bf CL}(1)_\om$ is  a strict subclass of the class ${\bf CFL}_\om$ of context free \ol s
accepted by  pushdown B\"uchi automata. 

If we omit the counter of a real-time B\"uchi $1$-counter automaton, then we simply get the 
notion of B\"uchi automaton. The class of \ol s accepted by  B\"uchi  automata  is  the 
class of regular \ol s.

\section{Hierarchies in a Cantor Space}

\subsection{Borel hierarchy and analytic sets}

 We assume the reader to be familiar with basic notions of topology which
may be found in \cite{Moschovakis80,LescowThomas,Staiger97,PerrinPin}.
There is a natural metric on the set $\Sio$ of  infinite words 
over a finite alphabet 
$\Si$ containing at least two letters which is called the {\it prefix metric} and is defined as follows. For $u, v \in \Sio$ and 
$u\neq v$ let $\delta(u, v)=2^{-l_{\mathrm{pref}(u,v)}}$ where $l_{\mathrm{pref}(u,v)}$ 
 is the first integer $n$
such that the $(n+1)^{st}$ letter of $u$ is different from the $(n+1)^{st}$ letter of $v$. 
This metric induces on $\Sio$ the usual  Cantor topology in which the {\it open subsets} of 
$\Sio$ are of the form $W\cdot \Si^\om$, for $W\subseteq \Sis$.
A set $L\subseteq \Si^\om$ is a {\it closed set} iff its complement $\Si^\om - L$ 
is an open set.

 Define now the {\it Borel Hierarchy} of subsets of $\Si^\om$:

\begin{Deff}
For a non-null countable ordinal $\alpha$, the classes ${\bf \Si}^0_\alpha$
 and ${\bf \Pi}^0_\alpha$ of the Borel Hierarchy on the topological space $\Si^\om$ 
are defined as follows:
\nl ${\bf \Si}^0_1$ is the class of open subsets of $\Si^\om$, 
 ${\bf \Pi}^0_1$ is the class of closed subsets of $\Si^\om$, 
\nl and for any countable ordinal $\alpha \geq 2$: 
\nl ${\bf \Si}^0_\alpha$ is the class of countable unions of subsets of $\Si^\om$ in 
$\bigcup_{\gamma <\alpha}{\bf \Pi}^0_\gamma$.
 \nl ${\bf \Pi}^0_\alpha$ is the class of countable intersections of subsets of $\Si^\om$ in 
$\bigcup_{\gamma <\alpha}{\bf \Si}^0_\gamma$.
\end{Deff}

\noi 
 The class of 
{\it Borel\ sets} is $\borel\! :=\!\bigcup_{\xi <\omega_1}\ \borapxi\! =\!
\bigcup_{\xi <\omega_1}\ \bormpxi$, where $\om_1$ is the first uncountable ordinal.  
There are also some subsets of $\Si^\om$ which are not Borel. 
In particular 
the class of Borel subsets of $\Si^\om$ is strictly included into 
the class  ${\bf \Si}^1_1$ of {\it analytic sets} which are 
obtained by projection of Borel sets. 

\begin{Deff} 
A subset $A$ of  $\Si^\om$ is in the class ${\bf \Si}^1_1$ of {\it analytic} sets
iff there exists another finite set $Y$ and a Borel subset $B$  of  $(\Si \times Y)^\om$ 
such that $ x \in A \lra \exists y \in Y^\om $ such that $(x, y) \in B$, 
where $(x, y)$ is the infinite word over the alphabet $\Si \times Y$ such that
$(x, y)(i)=(x(i),y(i))$ for each  integer $i\geq 1$.
\end{Deff} 

   We now define completeness with regard to reduction by continuous functions. 
For a countable ordinal  $\alpha\geq 1$, a set $F\subseteq \Si^\om$ is said to be 
a ${\bf \Si}^0_\alpha$  
(respectively,  ${\bf \Pi}^0_\alpha$, ${\bf \Si}^1_1$)-{\it complete set} 
iff for any set $E\subseteq Y^\om$  (with $Y$ a finite alphabet): 
 $E\in {\bf \Si}^0_\alpha$ (respectively,  $E\in {\bf \Pi}^0_\alpha$,  $E\in {\bf \Si}^1_1$) 
iff there exists a continuous function $f: Y^\om \ra \Si^\om$ such that $E = f^{-1}(F)$. 

 Let us now  recall the definition of the  arithmetical hierarchy of  \ol s, 
see for example \cite{Staiger97,Moschovakis80}.
 Let $\Si$ be a finite alphabet. An \ol~ $L\subseteq \Si^\om$  belongs to the class 
$\Si_n$ if and only if there exists a recursive relation 
$R_L\subseteq (\mathbb{N})^{n-1}\times \Si^\star$  such that
$$L = \{\sigma \in \Si^\om \mid \exists a_1\ldots Q_na_n  \quad (a_1,\ldots , a_{n-1}, 
\sigma[a_n+1])\in R_L \}$$
\noi where $Q_i$ is one of the quantifiers $\fa$ or $\exists$ 
(not necessarily in an alternating order). An \ol~ $L\subseteq \Si^\om$  belongs to the class 
$\Pi_n$ if and only if its complement $\Si^\om - L$  belongs to the class 
$\Si_n$.  
The inclusion relations that hold  between the classes $\Si_n$ and $\Pi_n$ are 
the same as for the corresponding classes of the Borel hierarchy and 
the classes $\Si_n$ and $\Pi_n$ are strictly included in the respective classes 
${\bf \Si}_n^0$ and ${\bf \Pi}_n^0$ of the Borel hierarchy. 

As in the case of the Borel hierarchy, projections of arithmetical sets 
(of the second $\Pi$-class) lead 
beyond the arithmetical hierarchy, to the analytical hierarchy of \ol s. 
 The first class of the analytical hierarchy of \ol s
 is the (lightface)  class $\Si^1_1$ of effective analytic sets.  
An \ol~ $L\subseteq \Si^\om$  belongs to the class 
$\Si_1^1$ if and only if there exists a recursive relation 
$R_L\subseteq (\mathbb{N})\times \{0, 1\}^\star \times \Si^\star$  such that:
$$L = \{\sigma \in \Si^\om \mid \exists \tau (\tau\in \{0, 1\}^\om \wedge \fa n \exists m 
 ( (n, \tau[m], \sigma[m]) \in R_L )) \}$$
\noi Thus an \ol~ $L\subseteq \Si^\om$  is in the class $\Si_1^1$ iff it is the projection 
of an \ol~ over the alphabet $\{0, 1\} \times \Si$ which is in the class $\Pi_2$  
 of the arithmetical hierarchy.

Kechris, Marker and Sami proved in \cite{KMS89} that the supremum 
of the set of Borel ranks of  (lightface) $\Pi_1^1$, so also of  (lightface) $\Si_1^1$,  sets is the ordinal $\gamma_2^1$. 
This ordinal is precisely defined in \cite{KMS89}. 
It holds that $ \om_1^{\mathrm{CK}} < \gamma_2^1$, where $\om_1^{\mathrm{CK}}$ is the first non-recursive ordinal,  called the Chruch-Kleene ordinal.  But 
the exact value of the ordinal $\gamma_2^1$ may depend on axioms of  set theory  \cite{KMS89}.
 
Notice  that it seems still unknown  whether {\it every } non null ordinal $\gamma < \gamma_2^1$ is the Borel rank 
of a (lightface) $\Pi_1^1$ (or $\Si_1^1$) set. 
On the other hand it is known that every ordinal $\gamma < \om_1^{\mathrm{CK}}$ is the Borel rank 
of a (lightface) $\Delta_1^1$-set, since for every ordinal 
$\gamma < \om_1^{\mathrm{CK}}$ 
there exist some 
${\bf \Si}^0_\gamma$-complete and   ${\bf \Pi}^0_\gamma$-complete 
sets in the class $\Delta_1^1$.  

  Recall that a B\"uchi Turing machine is just a Turing machine working on infinite 
inputs with a B\"uchi-like acceptance condition, and 
that the class of  $\om$-languages accepted by  B\"uchi Turing machines 
is the class $ \Si^1_1$ of effective analytic sets  \cite{CG78b,Staiger97}.

\subsection{Wadge hierarchy}

\noi We now introduce the Wadge hierarchy, which is a great refinement of the Borel hierarchy defined 
via reductions by continuous functions, \cite{Duparc01,Wadge83}. 

\begin{Deff}[Wadge \cite{Wadge83}] Let $X$, $Y$ be two finite alphabets. 
For $L\subseteq X^\om$ and $L'\subseteq Y^\om$, $L$ is said to be Wadge reducible to $L'$
($L\leq _W L')$ iff there exists a continuous function $f: X^\om \ra Y^\om$, such that
$L=f^{-1}(L')$.
l $L$ and $L'$ are Wadge equivalent iff $L\leq _W L'$ and $L'\leq _W L$. 
This will be denoted by $L\equiv_W L'$. And we shall say that 
$L<_W L'$ iff $L\leq _W L'$ but not $L'\leq _W L$.
\nl  A set $L\subseteq X^\om$ is said to be self dual iff  $L\equiv_W L^-$, and otherwise 
it is said to be non self dual.
\end{Deff}

\noi
 The relation $\leq _W $  is reflexive and transitive,
 and $\equiv_W $ is an equivalence relation.
 The {\it equivalence classes} of $\equiv_W $ are called {\it Wadge degrees}.  The Wadge hierarchy $WH$ is the class of Borel subsets of a set  $X^\om$, where  $X$ is a finite set,
 equipped with $\leq _W $ and with $\equiv_W $.
\nl  For $L\subseteq X^\om$ and $L'\subseteq Y^\om$, if   
$L\leq _W L'$ and $L=f^{-1}(L')$  where $f$ is a continuous 
function from $ X^\om$  into $Y^\om$, then $f$ is called a continuous reduction of $L$ to 
$L'$. Intuitively it means that $L$ is less complicated than $L'$ because 
to check whether $x\in L$ it suffices to check whether $f(x)\in L'$ where $f$ 
is a continuous function. Hence the Wadge degree of an \ol~
is a measure 
of its topological complexity. 
\nl
Notice  that in the above definition, we consider that a subset $L\subseteq  X^\om$ is given
together with the alphabet $X$.

\noi We can now define the {\it Wadge class} of a set $L$:

\begin{Deff}
Let $L$ be a subset of $X^\om$. The Wadge class of $L$ is :
\nl ~~~~~~~~~~~~~~~~~~ $[L]= \{ L' \mid  L'\subseteq Y^\om \mbox{ for a finite alphabet }Y   \mbox{  and  } L'\leq _W L \}.$ 
\end{Deff}

\noi Recall that each {\it Borel class} ${\bf \Si^0_\alpha}$ and ${\bf \Pi^0_\alpha}$ 
is a {\it Wadge class}. 
A set $L\subseteq X^\om$ is a ${\bf \Si^0_\alpha}$
 (respectively ${\bf \Pi^0_\alpha}$)-{\it complete set} iff for any set 
$L'\subseteq Y^\om$, $L'$ is in 
${\bf \Si^0_\alpha}$ (respectively ${\bf \Pi^0_\alpha}$) iff $L'\leq _W L . $
 
  There is a close relationship between Wadge reducibility
 and games which we now introduce.  

\begin{Deff} Let 
$L\subseteq X^\om$ and $L'\subseteq Y^\om$. 
The Wadge game  $W(L, L')$ is a game with perfect information between two players,
player 1 who is in charge of $L$ and player 2 who is in charge of $L'$.
Player 1 first writes a letter $a_1\in X$, then player 2 writes a letter
$b_1\in Y$, then player 1 writes a letter $a_2\in  X$, and so on. 
 The two players alternatively write letters $a_n$ of $X$ for player 1 and $b_n$ of $Y$
for player 2.
After $\om$ steps, the player 1 has written an $\om$-word $a\in X^\om$ and the player 2
has written an $\om$-word $b\in Y^\om$.
 The player 2 is allowed to skip, even infinitely often, provided he really writes an
$\om$-word in  $\om$ steps. The player 2 wins the play iff [$a\in L \lra b\in L'$], i.e. iff : 
\begin{center}
  [($a\in L ~{\rm and} ~ b\in L'$)~ {\rm or} ~ 
($a\notin L ~{\rm and}~ b\notin L'~{\rm and} ~ b~{\rm is~infinite}  $)].
\end{center}
\end{Deff}

\noi
Recall that a strategy for player 1 is a function 
$\sigma :(Y\cup \{s\})^\star\ra X$.
And a strategy for player 2 is a function $f:X^+\ra Y\cup\{ s\}$.
\nl $\sigma$ is a winning stategy  for player 1 iff he always wins a play when
 he uses the strategy $\sigma$, i.e. when the  $n^{th}$  letter he writes is given
by $a_n=\sigma (b_1\cdots b_{n-1})$, where $b_i$ is the letter written by player 2 
at step $i$ and $b_i=s$ if player 2 skips at step $i$.
\nl A winning strategy for player 2 is defined in a similar manner.

      Martin's Theorem states that every Gale-Stewart game $G(X)$ (see \cite{Kechris94}),  with $X$ a Borel set, 
is determined and this implies the following :

\begin{The} [Wadge] Let $L\subseteq X^\om$ and $L'\subseteq Y^\om$ be two Borel sets, where
$X$ and $Y$ are finite  alphabets. Then the Wadge game $W(L, L')$ is determined:
one of the two players has a winning strategy. And $L\leq_W L'$ iff the player 2 has a 
winning strategy  in the game $W(L, L')$.
\end{The}

\begin{The} [Wadge]\label{wh}
Up to the complement and $\equiv _W$, the class of Borel subsets of $X^\om$,
 for  a finite alphabet $X$  having at least two letters, is a well ordered hierarchy.
 There is an ordinal $|WH|$, called the length of the hierarchy, and a map
$d_W^0$ from $WH$ onto $|WH|-\{0\}$, such that for all $L, L' \subseteq X^\om$:
\nl $d_W^0 L < d_W^0 L' \lra L<_W L' $  and 
\nl $d_W^0 L = d_W^0 L' \lra [ L\equiv_W L' $ or $L\equiv_W L'^-]$.
\end{The}

\noi 
 The Wadge hierarchy of Borel sets of {\bf finite rank }
has  length $^1\varepsilon_0$ where $^1\varepsilon_0$
 is the limit of the ordinals $\alpha_n$ defined by $\alpha_1=\om_1$ and 
$\alpha_{n+1}=\om_1^{\alpha_n}$ for $n$ a non negative integer, $\om_1$
 being the first non countable ordinal. Then $^1\varepsilon_0$ is the first fixed 
point of the ordinal exponentiation of base $\om_1$. The length of the Wadge hierarchy 
of Borel sets in ${\bf \Delta^0_\om}= {\bf \Si^0_\om}\cap {\bf \Pi^0_\om}$ 
  is the $\om_1^{th}$ fixed point 
of the ordinal exponentiation of base $\om_1$, which is a much larger ordinal. The length 
of the whole Wadge hierarchy of Borel sets is a huge ordinal, with regard 
to the $\om_1^{th}$ fixed point 
of the ordinal exponentiation of base $\om_1$. It is described in \cite{Wadge83,Duparc01} 
by the use of the Veblen functions.

\section{Wadge Degrees of $\om$-Languages of Petri Nets }

We are firstly going to prove the following result. 

\begin{The}\label{thewad}    
The Wadge hierarchy of the class {\bf r}-${\bf BCL}(4)_\om$ is equal to the Wadge hierarchy of the class 
{\bf r}-${\bf CL}(1)_\om$. 
\end{The}

 In order to prove this result, we  first define 
 a coding of $\om$-words 
over a finite alphabet $\Si$ by $\om$-words over the alphabet  $\Si\cup\{A, B, 0\}$ where  
$A$, $B$ and $0$  are  new letters not in $\Si$.

We shall code an $\om$-word $x\in \Si^{\om}$ by the $\om$-word $h(x)$ defined by
$$h(x)=A0x(1)B0^{2}x(2)A\cdots      B0^{2n}x(2n)A0^{2n+1}x(2n+1)B \cdots  $$

\noi   This coding defines a  mapping  $h: \Si^{\om} \ra (\Si\cup\{A, B, 0\})^\om$. 
\nl The 
function $h$ is continuous because for all $\om$-words $x, y \in \Si^{\om}$ and 
each positive integer $n$,  it holds 
that  $\delta(x, y) < 2^{-n} \ra \delta( h(x), h(y) ) < 2^{-n}$. 

We now state the following lemma. 

\begin{lem}\label{lem5}  Let $\mathcal{A}$ be a real time 
$1$-counter B\"uchi automaton accepting 
$\om$-words over the alphabet $\Si$. Then one can construct a real time 4-blind-counter B\"uchi automaton $\mathcal{B}$ 
reading words over the alphabet $\Ga=\Si \cup\{A, B, 0\}$, such that
$ L(\mathcal{A})$ = $h^{-1} (L(\mathcal{B}))$, i.e.

  $$\fa x \in \Si^{\om} ~~~~ h(x) \in L(\mathcal{B}) \longleftrightarrow 
x\in  L(\mathcal{A}).$$ 
\end{lem}

 \proo
Let $\mathcal{A}$$=(K,\Si, \Delta, q_0, F)$ be a real time 
$1$-counter B\"uchi automaton accepting 
$\om$-words over the alphabet $\Si$. We are going to explain informally the behaviour of the 4-blind-counter B\"uchi automaton $\mathcal{B}$ 
when reading an $\om$-word of the form $h(x)$, even if we are going to  see  that $\mathcal{B}$ may also accept some infinite words which do not belong to the 
range of $h$. Recall that $h(x)$ is of the form
$$h(x)=A0x(1)B0^{2}x(2)A \cdots  
B0^{2n}x(2n)A0^{2n+1}x(2n+1)B \cdots  $$

\noi Notice that in particular every $\om$-word in $h(\Sio)$ is of the form: 
$$ y = A0^{n_1}x(1)B0^{n_2}x(2)A \cdots  
 B0^{n_{2n}}x(2n)A0^{n_{2n+1}}x(2n+1)B \cdots  $$

\noi where for all $i\geq 1$,  $n_i >0$ is a  positive integer, and $x(i)\in \Si$. 

\hs Moreover it is easy to see that the set of $\om$-words $y \in \Gao$  which can be written in the  above form is a regular $\om$-language $\mathcal{R}\subseteq \Gao$, and thus we can assume, using a classical product construction (see for instance \cite{PerrinPin}),  that the automaton 
$\mathcal{B}$ will only  accept some $\om$-words of this form. 

 Now the  reading by the automaton $\mathcal{B}$  of  an $\om$-word of  the above form
$$ y = A0^{n_1}x(1)B0^{n_2}x(2)A \cdots   B0^{n_{2n}}x(2n)A0^{n_{2n+1}}x(2n+1)B \cdots  $$
 
  \noi  will give a  decomposition of the $\om$-word $y$ of the following form: 
$$y = Au_1v_1x(1)Bu_2v_2x(2)Au_3v_3x(3)B \cdots   Bu_{2n}v_{2n}x(2n)Au_{2n+1}v_{2n+1}x(2n+1)B \cdots  $$

\noi where,  for all integers $i\geq 1$, $ u_i, v_i  \in 0^\star$, $x(i) \in \Si$, $|u_1|=0$. 

 The automaton  $\mathcal{B}$  will  use its four {\it blind} counters, which we denote $\mathcal{C}_1,  \mathcal{C}_2,  \mathcal{C}_3,  \mathcal{C}_4,$  in the following way. 
Recall that the automaton  $\mathcal{B}$  being non-deterministic, we do not describe the unique run of $\mathcal{B}$ on $y$, but the general case of a possible run. 

At the beginning of the run, the value of each of the four counters is equal to zero. Then  the counter  $\mathcal{C}_1$ is increased of $|u_1|$ when reading $u_1$, i.e. the counter  $\mathcal{C}_1$ is actually not increased since $|u_1|=0$ and the finite control is here used to check this. 
Then the counter $\mathcal{C}_2$ is increased of $1$ for each letter $0$ of $v_1$ which is read until the automaton  reads the letter $x(1)$ and then the letter $B$. Notice that at this time the values of the counters $\mathcal{C}_3$ and   $\mathcal{C}_4$ are still equal to zero. 
Then the behaviour of the automaton $\mathcal{B}$ when reading the next segment $0^{n_2}x(2)A$ is as follows. The counters 
 $\mathcal{C}_1$ is firstly decreased of $1$ for each letter $0$ read,  when reading $k_2$ letters $0$, where $k_2\geq 0$ (notice that here $k_2=0$ because the value of the counter 
$\mathcal{C}_1$ being equal to zero, it cannot decrease under $0$).  Then   the counter $\mathcal{C}_2$ is decreased of $1$ for each letter $0$ read, and next the automaton has to read one more letter $0$, leaving unchanged the counters $\mathcal{C}_1$ and   $\mathcal{C}_2$, before reading the letter $x(2)$.  The end of the decreasing mode of $\mathcal{C}_1$ coincide with the beginning of the decreasing mode of $\mathcal{C}_2$, and this change may occur in a {\it non-deterministic way} (because the automaton $\mathcal{B}$ cannot check whether the value of $\mathcal{C}_1$ is equal to zero).
 Now we describe the behaviour of the counters $\mathcal{C}_3$ and   $\mathcal{C}_4$ when reading the segment $0^{n_2}x(2)A$. Using its finite control,  the automaton $\mathcal{B}$
 has checked that $|u_1|=0$, and then  if there is a transition of 
the automaton  $\mathcal{A}$ such that $x(1) : ( q_{0}, |u_1|) \mapsto_{\mathcal{A}} 
(q_1, |u_1| +N_1 )$ then the counter $\mathcal{C}_3$ is increased of $1$ for each letter $0$ read, during the reading of the $k_2+N_1$ first  letters $0$ of $0^{n_2}$, where $k_2$ is described above as the number of which the counter $\mathcal{C}_1$ has been decreased. This determines $u_2$ by $|u_2|=k_2+N_1$ and then the counter $\mathcal{C}_4$ is increased by $1$ for each letter $0$ read until $\mathcal{B}$ reads $x(2)$, and this determines $v_2$.  Notice that the automaton $\mathcal{B}$ keeps in its finite control the memory of the state $q_1$ of the automaton $\mathcal{A}$, and  that, after having read the segment $0^{n_2}=u_2v_2$, the values of the counters $\mathcal{C}_3$ and   $\mathcal{C}_4$ are respectively 
$|\mathcal{C}_3|=|u_2|=k_2+N_1$ and $|\mathcal{C}_4|=|v_2|=n_2-(|u_2|)$. 

Now the run will continue. Notice that generally when reading a segment $B0^{n_{2n}}x(2n)A$  the counters $\mathcal{C}_1$ and $\mathcal{C}_2$ will  successively decrease when reading 
the first $(n_{2n}-1)$ letters $0$ and then will remain unchanged when reading the last letter $0$, and the counters $\mathcal{C}_3$ and   $\mathcal{C}_4$ will successively increase, when reading the $(n_{2n})$ letters $0$. Again the end of the decreasing mode of $\mathcal{C}_1$ coincide with the beginning of the decreasing mode of $\mathcal{C}_2$, and this change may occur in a {\it non-deterministic way}.  But the automaton has kept in its finite control 
whether $|u_{2n-1}|=0$ or not and also a state $q_{2n-2}$ of the automaton  $\mathcal{A}$. Now,  if  there is a transition of 
the automaton  $\mathcal{A}$ such that $x(2n-1) : ( q_{2n-2}, |u_{2n-1}|) \mapsto_{\mathcal{A}} 
(q_{2n-1}, |u_{2n-1}| +N_{2n-1} )$ for some integer $N_{2n-1} \in \{-1; 0, 1\}$, and the counter     $\mathcal{C}_1$ is decreased of $1$ for each letter $0$ read,  when reading $k_{2n}$ first letters $0$ of $0^{n_{2n}}$, then the counter  $\mathcal{C}_3$ is increased of $1$ for each letter $0$ read, during the reading of the $k_{2n}+N_{2n-1}$ first  letters $0$ of $0^{n_{2n}}$, and next the   counter $\mathcal{C}_4$  is increased by $1$ for each letter $0$ read until $\mathcal{B}$ reads $x(2n)$, and this determines $v_{2n}$.  Then after having read the segment $0^{n_ {2n}}=u_{2n}v_{2n}$, the values of the counters $\mathcal{C}_3$ and   $\mathcal{C}_4$ have respectively increased of $|u_{2n}|=k_{2n}+N_{2n-1}$ and $|v_{2n}|=n_{2n}-|u_{2n}|$. 
Notice that one cannot ensure that, after the reading of $0^{n_ {2n}}=u_{2n}v_{2n}$,  the exact values of these counters  are   $|\mathcal{C}_3|=|u_{2n}|=k_{2n}+N_{2n-1}$ and $|\mathcal{C}_4|=|v_{2n}| =n_ {2n} - |u_{2n}|$. Actually this is due to the fact that one cannot ensure that the values of $\mathcal{C}_3$ and   $\mathcal{C}_4$ are equal to zero at the beginning of the reading of the segment $B0^{n_{2n}}x(2n)A$ although we will see this is true and important in the particular case of a word of the form $y=h(x)$. 

The run will continue in a similar manner during the reading of the next segment 	$A0^{n_{2n+1}}x(2n+1)B$, but here the role of the counters $\mathcal{C}_1$ and $\mathcal{C}_2$ on one side, and of the counters $\mathcal{C}_3$ and   $\mathcal{C}_4$  on the other side, will be interchanged.  More precisely the counters $\mathcal{C}_3$ and $\mathcal{C}_4$ will  successively decrease when reading 
the first $(n_{2n+1}-1)$ letters $0$ and then will remain unchanged when reading the last letter $0$, and the counters $\mathcal{C}_1$ and   $\mathcal{C}_2$ will successively increase, when reading the $(n_{2n+1})$ letters $0$. The end of the decreasing mode of $\mathcal{C}_3$ coincide with the beginning of the decreasing mode of $\mathcal{C}_4$, and this change may occur in a {\it non-deterministic way}.  But the automaton has kept in its finite control 
whether $|u_{2n}|=0$ or not and also a state $q_{2n-1}$ of the automaton  $\mathcal{A}$. Now,  if  there is a transition of 
the automaton  $\mathcal{A}$ such that $x(2n) : ( q_{2n-1}, |u_{2n}|) \mapsto_{\mathcal{A}} 
(q_{2n}, |u_{2n}| +N_{2n} )$ for some integer $N_{2n} \in \{-1; 0, 1\}$, and the counter     $\mathcal{C}_3$ is decreased of $1$ for each letter $0$ read,  when reading $k_{2n+1}$ first letters $0$ of $0^{n_{2n+1}}$, then the counter  $\mathcal{C}_1$ is increased of $1$ for each letter $0$ read, during the reading of the $k_{2n+1}+N_{2n}$ first  letters $0$ of $0^{n_{2n+1}}$, and next the   counter $\mathcal{C}_2$  is increased by $1$ for each letter $0$ read until $\mathcal{B}$ reads $x(2n+1)$, and this determines $v_{2n+1}$.  Then after having read the segment $0^{n_ {2n+1}}=u_{2n+1}v_{2n+1}$, the values of the counters $\mathcal{C}_1$ and   $\mathcal{C}_2$ have respectively increased of $|u_{2n+1}|=k_{2n+1}+N_{2n}$ and $|v_{2n+1}|=n_{2n+1}-|u_{2n+1}|$. 
Notice that again one cannot ensure that, after the reading of $0^{n_ {2n+1}}=u_{2n+1}v_{2n+1}$,  the exact values of these counters  are   $|\mathcal{C}_1|=|u_{2n+1}|=k_{2n+1}+N_{2n}$ and $|\mathcal{C}_2|=|v_{2n+1}| =n_ {2n+1} - |u_{2n+1}|$. This is due to the fact that one cannot ensure that the values of $\mathcal{C}_1$ and   $\mathcal{C}_2$ are equal to zero at the beginning of the reading of the segment $A0^{n_{2n+1}}x(2n+1)B$ although we will see this is true and important in the particular case of a word of the form $y=h(x)$. 

The run then continues in the same way if it is possible and in particular if there is no blocking due to the fact that one of the counters of  the automaton  $\mathcal{B}$ would have a negative value. 

Now an $\om$-word $y\in \mathcal{R}\subseteq \Gao$ of the above form will be accepted by the automaton  $\mathcal{B}$ if there is such an infinite  run for which a final state $q_f\in F$ 
of the automaton  $\mathcal{A}$ has been stored infinitely often in the finite control of $\mathcal{B}$ in the way which has just been described above. 

 We now consider the particular case of an $\om$-word of the form $y=h(x)$,  for some  $x\in \Sio$. 

 Let then
 $y=h(x)=A0x(1)B0^{2}x(2)A0^{3}x(3)B \cdots B0^{2n}x(2n)A0^{2n+1}x(2n+1)B \cdots $

 We are going to show that, if  $y$ is accepted by the automaton $\mathcal{B}$, then $x\in L(\mathcal{A})$. Let us consider a run of the automaton $\mathcal{B}$ on $y$ as described above and which is an accepting run. We first show by induction on $n\geq 1$, that after having read an initial segment 
of the form $$A0x(1)B0^{2}x(2)A\cdots A0^{2n-1}x(2n-1)B$$ 
\noi the values of the counters $\mathcal{C}_3$ and   $\mathcal{C}_4$ are equal to zero, and the values of the 
counters $\mathcal{C}_1$ and   $\mathcal{C}_2$ satisfy $|\mathcal{C}_1| + |\mathcal{C}_2|=2n-1$. And similarly after having read an initial segment 
of the form $A0x(1)B0^{2}x(2)A \cdots B0^{2n}x(2n)A$ the values of the counters $\mathcal{C}_1$ and   $\mathcal{C}_2$ are equal to zero, and the values of the 
counters $\mathcal{C}_3$ and   $\mathcal{C}_4$ satisfy $|\mathcal{C}_3| + |\mathcal{C}_4|=2n$. 

For $n=1$, we have seen that after having read the initial segment $A0x(1)B$, the values of the counters $\mathcal{C}_1$ and   $\mathcal{C}_2$ will be respectively $0$ and $|v_1|$ and here $|v_1|=1$ and thus $|\mathcal{C}_1| + |\mathcal{C}_2|=1$. On the other hand the counters  $\mathcal{C}_3$ and   $\mathcal{C}_4$  have not yet increased so that the value of each of these counters is equal to zero. During  the reading of the segment $0^{2}$ of  $0^{2}x(2)A$ the counters $\mathcal{C}_1$ and   $\mathcal{C}_2$  successively decrease. But here 
 $\mathcal{C}_1$ cannot decrease  (with the above notations, it holds that $k_2=0$) so $\mathcal{C}_2$ must decrease of $1$   because after the decreasing mode the automaton $\mathcal{B}$ must read a last letter $0$ without decreasing 
 the counters  $\mathcal{C}_1$ and   $\mathcal{C}_2$ and then the letter $x(2)\in \Si$. Thus after having read $0^{2}x(2)A$ the values of $\mathcal{C}_1$ and   $\mathcal{C}_2$ are equal to zero. Moreover the counters $\mathcal{C}_3$ and   $\mathcal{C}_4$ had their values equal to zero at the beginning of the reading of $0^{2}x(2)A$ and they successively increase during the reading of $0^{2}$ and they remain unchanged during the reading of $x(2)A$ so that their values satisfy  $|\mathcal{C}_3| + |\mathcal{C}_4|=2$ after the reading of 
  $0^{2}x(2)A$. 
  
  Assume now that for some integer $n>1$  the claim is proved for all integers $k<n$ and let us prove it for the integer $n$. By induction hypothesis we know that at the beginning of the reading of the segment 
  $A0^{2n-1}x(2n-1)B$ of $y$, the  values of the counters $\mathcal{C}_1$ and   $\mathcal{C}_2$ are equal to zero, and the values of the 
counters $\mathcal{C}_3$ and   $\mathcal{C}_4$ satisfy $|\mathcal{C}_3| + |\mathcal{C}_4|=2n-2$. When reading the $(2n-2)$ first letters $0$ of $A0^{2n-1}x(2n-1)B$ the 
counters $\mathcal{C}_3$ and   $\mathcal{C}_4$ successively decrease and they must decrease completely because after there must remain only one letter $0$ to be read by 
$\mathcal{B}$ before the letter $x(2n-1)$. Therefore after the reading of  $A0^{2n-1}x(2n-1)B$  the values of the counters $\mathcal{C}_3$ and   $\mathcal{C}_4$ are equal to zero. 
And since the   values of the counters $\mathcal{C}_1$ and   $\mathcal{C}_2$ are equal to zero before the reading of $0^{2n-1}x(2n-1)B$ and these counters  successively increase during the reading of  $0^{2n-1}$, their values satisfy $|\mathcal{C}_1| + |\mathcal{C}_2|=2n-1$ after the reading of $A0^{2n-1}x(2n-1)B$. 
We can reason in a very similar manner for the reading of the next segment  $B0^{2n}x(2n)A$, the role  of the counters $\mathcal{C}_1$ and $\mathcal{C}_2$ on one side, and of the counters $\mathcal{C}_3$ and   $\mathcal{C}_4$  on the other side, being simply  interchanged.  This  ends the proof of the claim by induction on $n$. 
  
  It is now easy to see by induction that for each integer $n\geq 2$, it holds that $k_n=|u_{n-1}|$.  Then , since with the above notations we have  
  $|u_{n+1}|=k_{n+1} + N_{n}=|u_{n}| + N_n$, and there is a transition of 
the automaton  $\mathcal{A}$ such that $x(n) : ( q_{n-1}, |u_{n}|) \mapsto_{\mathcal{A}} 
(q_{n}, |u_{n}| +N_{n})$ for $N_{n} \in \{-1; 0, 1\}$, it holds that  $x(n) : ( q_{n-1}, |u_{n}|) \mapsto_{\mathcal{A}} 
(q_{n}, |u_{n+1}| )$. Therefore the sequence $(q_i, |u_{i}|)_{i\geq 0}$ is an accepting run of the automaton $\mathcal{A}$ on the $\om$-word $x$ and $x\in L(\mathcal{A})$. 
 Notice that the state $q_0$ of the sequence $(q_i)_{i\geq 0}$  is also the initial state 
of $\mathcal{A}$. 
  
  Conversely, it is easy to see that if $x\in L(\mathcal{A})$ then there exists an accepting run of the automaton $\mathcal{B}$ on the $\om$-word $h(x)$ and $h(x)\in L(\mathcal{B})$. 
  \ep

\hs  The above Lemma \ref{lem5} shows that, given  a real time 
$1$-counter (with zero-test) B\"uchi automaton $\mathcal{A}$ accepting 
$\om$-words over the alphabet $\Si$,  one can construct a real time 4-blind-counter B\"uchi automaton $\mathcal{B}$ which can  simulate the $1$-counter automaton  $\mathcal{A}$  on the code $h(x)$ of the word $x$. On the other hand, we cannot describe precisely the $\om$-words which are accepted by $\mathcal{B}$ but are not in the set  $h(\Sio)$. However we can see 
that all these words have a special shape, as stated by the following lemma. 

\begin{lem}\label{lem6}  Let $\mathcal{A}$ be a real time 
$1$-counter B\"uchi automaton accepting 
$\om$-words over the alphabet $\Si$,  and let    $\mathcal{B}$   be the real time 4-blind-counter B\"uchi automaton 
reading words over the alphabet $\Ga=\Si \cup\{A, B, 0\}$ which is constructed in the proof of Lemma \ref{lem5}. 
Let $y \in L(\mathcal{B})\setminus h(\Sio)$ being of the following form
$$ y = A0^{n_1}x(1)B0^{n_2}x(2)A0^{n_3}x(3)B \cdots  B0^{n_{2n}}x(2n)A0^{n_{2n+1}}x(2n+1)B \cdots  $$
\noi and let $i_0$ be the smallest integer $i$ such that $n_i \neq i$. Then it holds that either $i_0=1$ or $n_{i_0} < i_0$. 
\end{lem}

  Let $\mathcal{L} \subseteq \Gao$ be the $\om$-language containing the $\om$-words over $\Gamma$   which belong to one of the following $\om$-languages. 

\begin{itemize} 
\ite $\mathcal{L}$$_1$ is the set of $\om$-words over the alphabet $\Si\cup\{A, B, 0\}$ 
which have not any initial segment in  $A\cdot 0\cdot \Si \cdot  B$. 

\ite $\mathcal{L}$$_2$ is the set of $\om$-words over the alphabet $\Si\cup\{A, B, 0\}$ 
which contain a segment of the form  $B\cdot 0^n \cdot a \cdot A\cdot 0^m\cdot b$ or of the form 
$A \cdot 0^n \cdot a \cdot B \cdot 0^m\cdot b$
for some letters $a, b \in \Si$ and some positive integers $m \leq  n$. 
\end{itemize}

\begin{lem}\label{lem7}
The $\om$-language $\mathcal{L}$ is accepted by a (non-deterministic)  real-time $1$-blind counter B\"uchi  automaton.  
\end{lem}

\proo
First, it is easy to see that 
$\mathcal{L}$$_1$ is in fact a regular $\om$-language, and thus it is also accepted by  a real-time $1$-blind counter B\"uchi  automaton (even without active counter).  On the other hand it is also easy to construct a real time
$1$-blind counter B\"uchi  automaton accepting the $\om$-language $\mathcal{L}$$_2$. The class of $\om$-languages accepted by {\it non-deterministic} real time
$1$-blind counter B\"uchi  automata being closed under finite union in an effective way, one can  construct a real time $1$-blind counter B\"uchi  automaton accepting $\mathcal{L}$. 
\ep

\begin{lem}\label{lem8}
 Let $\mathcal{A}$ be a real time 
$1$-counter B\"uchi automaton accepting 
$\om$-words over the alphabet $\Si$. Then one can construct a  real time $4$-blind counter B\"uchi  automaton $\mathcal{P}_{\mathcal{A}}$ such that 
$$L(\mathcal{P}_{\mathcal{A}}) = h( L(\mathcal{A}) ) \cup \mathcal{L}.$$
\end{lem}

\proo 
 Let $\mathcal{A}$ be a real time 
$1$-counter B\"uchi automaton accepting 
$\om$-words over $\Si$. We have seen in the proof of Lemma \ref{lem5} that one can construct a real time $4$-blind counter B\"uchi  automaton 
$\mathcal{B}$ 
reading words over the alphabet $\Ga=\Si \cup\{A, B, 0\}$, such that
$ L(\mathcal{A})$ = $h^{-1} (L(\mathcal{B}))$, i.e. 
$\fa x \in \Si^{\om} ~~~~ h(x) \in L(\mathcal{B}) $$\longleftrightarrow 
x\in  L(\mathcal{A})$. Moreover By Lemma \ref{lem6} it holds that $L(\mathcal{B})\setminus h(\Sio) \subseteq \mathcal{L}$. and thus 
$$h( L(\mathcal{A}) ) \cup \mathcal{L} = L(\mathcal{B})  \cup \mathcal{L}$$
\noi But By Lemma \ref{lem7} the $\om$-language $\mathcal{L}$ is accepted by a (non-deterministic)  real-time $1$-blind counter B\"uchi  automaton, hence also by a 
 real-time $4$-blind counter B\"uchi  automaton. The class of $\om$-languages  accepted by a (non-deterministic)  real-time $4$-blind counter B\"uchi  automata is closed under 
 finite union in an effective way, and thus  one can construct a  real time $4$-blind counter B\"uchi  automaton $\mathcal{P}_{\mathcal{A}}$ such that ~ ~
$L(\mathcal{P}_{\mathcal{A}}) = h( L(\mathcal{A}) ) \cup \mathcal{L}. $
\ep

\hs     We are now going to  prove that 
if $L(\mathcal{A})$$\subseteq \Sio$ is accepted by a 
real time $1$-counter 
automaton $\mathcal{A}$ with a B\"uchi acceptance condition then 
$L(\mathcal{P}_{\mathcal{A}}) = h( L(\mathcal{A}) )$$ \cup \mathcal{L}$ 
 will have the same Wadge degree as the $\om$-language 
$L(\mathcal{A})$, except for some very simple cases.

 We first notice that $h(\Si^{\om})$ is a closed subset of $\Gao$. Indeed  it is the image of the compact set $\Sio$ by the continuous function $h$, and thus it is a compact hence also closed subset of $\Gao = (\Si\cup\{A, B, 0\})^\om$.  Thus its complement 
$h(\Si^{\om})^-=(\Si\cup\{A, B, 0\})^\om - h(\Si^{\om})$ is an open subset of $\Gao$. Moreover  the set $\mathcal{L}$ is an open subset of  $\Gao$, as it can be easily seen from its definition and one can easily define, from the definition of the $\om$-language  $\mathcal{L}$, a finitary  language $V \subseteq \Gas$   such that  $\mathcal{L}=V\cdot \Gao$. 
We shall also denote $\mathcal{L}'=h(\Si^{\om})^-\setminus \mathcal{L}$ so that $\Gao$ is the dijoint union $\Gao = h(\Si^{\om}) \cup \mathcal{L}  \cup \mathcal{L}'$. Notice that 
$\mathcal{L}'$ is the difference of the  two open sets $h(\Si^{\om})^-$ and $\mathcal{L}$. 

\hs We now wish to return to the proof of the above Theorem \ref{thewad} stating that 
the Wadge hierarchy of the class {\bf r}-${\bf BCL}(4)_\om$  is equal to the Wadge hierarchy of the class   {\bf r}-${\bf CL}(1)_\om$.

    To prove this result we firstly consider non self dual Borel sets. We recall the definition of Wadge degrees 
introduced by Duparc in \cite{Duparc01} and which is a slight modification of the previous one. 

\begin{Deff}
\noi
\begin{enumerate}
\ite[(a)]  $d_w(\emptyset)=d_w(\emptyset^-)=1$
\ite[(b)]  $d_w(L)=sup \{d_w(L')+1 ~\mid ~~L' {\rm ~non~ self ~dual~ and~}
L'<_W L \} $
\nl (for either $L$ self dual or not, $L>_W \emptyset).$
\end{enumerate}
\end{Deff}

\noi  Wadge and Duparc used  the operation of sum of 
sets of infinite words which has as  
counterpart the ordinal
addition  over Wadge degrees.

\begin{Deff}[Wadge, see \cite{Wadge83,Duparc01}]
Assume that $X\subseteq Y$ are two finite alphabets,
  $Y-X$ containing at least two elements, and that
$\{X_+, X_-\}$ is a partition of $Y-X$ in two non empty sets.
 Let $L \subseteq X^{\om}$ and $L' \subseteq Y^{\om}$, then
 
 $L' + L =_{df} L\cup \{ u\cdot a\cdot \beta  ~\mid  ~ u\in X^\star , ~(a\in X_+
~and ~\beta \in L' )~
 or ~(a\in X_- ~and ~\beta \in L'^- )\}$
\end{Deff}

\noi This operation is closely related to the {\it ordinal sum}
 as it is stated in the following:

\begin{The}[Wadge, see \cite{Wadge83,Duparc01}]\label{thesum}
Let $X\subseteq Y$, $Y-X$ containing at least two elements,
   $L \subseteq X^{\om}$ and $L' \subseteq Y^{\om}$ be 
non self dual  Borel sets.
Then $(L+L')$ is a non self dual Borel set and
$d_w( L'+L )= d_w( L' ) + d_w( L )$.
\end{The}

\noi A player in charge of a set $L'+L$ in a Wadge game is like a player in charge of the set $L$ but who 
can, at any step of the play,    erase  his previous play and choose to be this time in charge of  $L'$ or of $L'^-$. 
Notice that he can do this only one time during a play. 

The following lemma was proved in \cite{Fin-mscs06}. Notice that below the emptyset is considered as an $\om$-language over an alphabet 
$\Delta$ such that $\Delta - \Si$ contains at least two elements. 

\begin{lem}\label{sum-wad}
Let $L \subseteq \Sio$ be a non self dual  Borel set such that $d_w( L )\geq \om$. Then it holds that $L \equiv_W \emptyset + L$. 
\end{lem}

We can now prove the following lemma. 

\begin{lem}\label{nad}
Let  $L \subseteq \Sio$ be a non self dual  Borel set acccepted by a real time 
$1$-counter B\"uchi automaton  $\mathcal{A}$. 
Then there is an $\om$-language $L'$ accepted by   a real time $4$-blind counter B\"uchi  automaton such that $L \equiv_W L'$.  
\end{lem}

\proo  Recall first  that there are regular $\om$-languages of every finite Wadge degree, \cite{Staiger97,Selivanov98}. These regular $\om$-languages 
are Boolean combinations of open sets, and they obviously  belong to the class  {\bf r}-${\bf BCL}(4)_\om$  since every regular $\om$-language belongs to this class.  

So we have only to consider the case of non self dual Borel sets of Wadge 
degrees greater than or equal to $\om$. 

 Let then $L=L(\mathcal{A}) \subseteq \Sio$ be a non self dual  Borel set, acccepted by a real time 
$1$-counter B\"uchi automaton  $\mathcal{A}$, such that $d_w( L )\geq \om$.   
By Lemma \ref{lem8},  
$L(\mathcal{P}_{\mathcal{A}}) = h( L(\mathcal{A}) )$$ \cup \mathcal{L}$ is accepted by a   a real time $4$-blind counter B\"uchi  automaton $\mathcal{P}_{\mathcal{A}}$, 
where the mapping $h: \Sio \ra (\Si \cup\{A, B, 0\})^\om$ is defined, for $x\in \Sio$,  by: 
$$h(x)= A0x(1)B0^{2}x(2)A0^{3}x(3)B \cdots    B0^{2n}x(2n)A0^{2n+1}x(2n+1)B \cdots $$

We set $L'=L(\mathcal{P}_{\mathcal{A}})$ and we now prove that  $L' \equiv_W L$.  

\hs Firstly, it is easy to see that $L \leq_W L'$. In order to prove this  we can consider the Wadge game 
$W( L, L' )$. It is easy to see that Player 2 has a winning strategy in this game which consists in essentially 
copying the play of Player 1, except that Player 2 actually writes a beginning of the code given by $h$ of what has been written by Player 1.  
This is achieved  in such a way that Player 2  has written the initial word 
$A0x(1)B0^{2}x(2)A \cdots  
B0^{2n}x(2n)$ while Player 1 has written the initial word $x(1)x(2) \cdots 
x(2n)$ (respectively, $A0x(1)B0^{2}x(2)A \cdots  
B0^{2n}x(2n)A0^{2n+1}x(2n+1)$ while Player 1 has written the initial word $x(1)x(2) \cdots 
x(2n+1)$)
Notice that one can admit that a player writes a finite word at each step of the play instead of a single letter. This does not change the 
winner of a Wadge game. At the end of a play if Player 1 has written the $\om$-word $x$ then Player 2 has written $h(x)$ and thus 
$x\in L(\mathcal{A}) \Longleftrightarrow h(x) \in L'$ and Player 2 wins the play.

\hs To prove that $L' \leq_W L $, it suffices to prove that $L' \leq_W \emptyset +(\emptyset + L)$ because 
Lemma \ref{sum-wad} states that $ \emptyset + L \equiv_W  L$, and thus also $\emptyset + (\emptyset + L) \equiv_W  L$. Consider the Wadge game $W( L',  \emptyset + (\emptyset + L) )$. 
Player 2 has a winning strategy in this game which we now describe. 

As long as Player 1 remains in the closed set $h(\Si^{\om})$ (this means that the word written by Player 1 is a prefix of some infinite word in $h(\Si^{\om})$) 
Player 2 essentially  copies  the play of player 1 except that Player 2 skips when player 1 writes a letter not in $\Si$. 
He continues forever with this strategy if the word written by player 1 is always a prefix of some $\om$-word of $h(\Sio)$. Then after $\om$ steps 
Player 1 has written an $\om$-word $h(x)$ for some $x \in \Sio$, and Player 2 has written $x$. So in that case 
$h(x) \in L'$ iff  $x \in L(\mathcal{A})$ iff  $x \in \emptyset + (\emptyset + L)$.  

But if at some step of the play, Player 1 ``goes out of" the closed set  $h(\Sio)$ because the word he has now 
written is not a prefix of any $\om$-word of  $h(\Sio)$,  then Player 1 ``enters'' in the open set $h(\Sio)^- = \mathcal{L}\cup \mathcal{L}'$ and will stay in this set. Two cases may now appear. 

{\bf First case.} When  Player 1 ``enters'' in the open set $h(\Sio)^- = \mathcal{L}\cup \mathcal{L}'$, he actually enters in the open set $ \mathcal{L}=V\cdot \Gao$ (this means that Player 1 has written an initial segment in $V$). Then the final word written by Player 1 will surely be inside $L'$. Player 2 can now  write a letter of $\Delta -\Si$ in such a way that he is now like a player in charge of the wholeset  and he can 
now writes an $\om$-word $u$ so that his final $\om$-word will be inside $\emptyset + L $, and also inside $\emptyset + (\emptyset + L)$. Thus Player 2 wins this play too. 

{\bf Second case.}   When  Player 1 ``enters'' in the open set $h(\Sio)^- = \mathcal{L}\cup \mathcal{L}'$, he does not enter in the open set $ \mathcal{L}=V\cdot \Gao$. 
Then  Player 2,  being first like a player in charge of the set $(\emptyset + L)$,  can write a  letter of $\Delta -\Si$ in such a way that he is now like a player in charge of the emptyset  and he can 
now continue, writing an $\om$-word $u$. If Player 1 never enters  in the open set $\mathcal{L}=V\cdot \Gao$ then the final word written by Player 1 will be in $\mathcal{L}'$ and thus surely outside $L'$, and the final word written by Player 2 will be outside the emptyset. So in that case Player 2 wins this play too. 
If at some step of the play Player  1 enters  in the open set $\mathcal{L}=V\cdot \Gao$ then his final $\om$-word will be surely in $L'$. In that case Player 1, in charge of the set 
$\emptyset + (\emptyset + L)$, can again write an extra letter and choose to be  in charge of the wholeset and he can 
now write an $\om$-word $v$ so that his final $\om$-word will be inside $\emptyset + (\emptyset + L)$. Thus Player 2 wins this play too. 

 Finally we have proved that $L \leq_W  L'   \leq_W L $ thus it holds that $L' \equiv_W L$.  This ends the proof. 
\nolinebreak \ep

\hs {\bf End of Proof of Theorem \ref{thewad}. } 
  
Let  $L \subseteq \Sio$ be a   Borel set accepted by a real time 
$1$-counter B\"uchi automaton  $\mathcal{A}$. 
If the Wadge degree of $L$ is finite, it is well known that it is Wadge 
equivalent to a regular $\om$-language, hence also to an  $\om$-language in  the class {\bf r}-${\bf BCL}(4)_\om$. 
If  $L$ is non self dual and its Wadge degree is greater than or equal to $\om$, then we know from Lemma \ref{nad} that 
there is an $\om$-language  $L'$ accepted by a   a real time $4$-blind counter B\"uchi  automaton such that $L \equiv_W L'$. 

 It remains to consider the case of self dual Borel sets.  
The alphabet $\Si$ being finite, a self dual Borel set $L$ is always Wadge equivalent to a Borel set 
in the form $\Si_1\cdot L_1 \cup \Si_2\cdot L_2$, where $(\Si_1, \Si_2)$ form a partition of $\Si$, 
and $L_1, L_2\subseteq \Sio$ are non self dual Borel sets such that 
$L_1 \equiv_W L_2^-$.  
Moreover $L_1$ and $L_2$ can be taken in the form $L_{(u_1)}=u_1\cdot \Sio \cap L$ and     
 $L_{(u_2)}=u_2\cdot \Sio \cap L$     for some $u_1, u_2 \in \Sis$, see
\cite{Duparc03}. 
So if  $L \subseteq \Sio$ is a self dual Borel set accepted by a real time 
$1$-counter B\"uchi automaton  
then $L \equiv_W \Si_1\cdot L_1 \cup \Si_2\cdot L_2$, where $(\Si_1, \Si_2)$ form a partition of $\Si$, and 
 $L_1, L_2\subseteq \Sio$ are non self dual Borel sets accepted by real time 
$1$-counter B\"uchi automata.  
We have already proved that there is  an     $\om$-language      $L'_1$  in  the class {\bf r}-${\bf BCL}(4)_\om$ such that $L'_1 \equiv_W L_1$ and 
 an $\om$-language $L'_2$ in  the class {\bf r}-${\bf BCL}(4)_\om$  such that $L_2'^-\equiv_W L_2$.  Thus 
$L  \equiv_W  \Si_1\cdot L_1 \cup \Si_2\cdot L_2  \equiv_W \Si_1\cdot L_1' \cup \Si_2\cdot L'_2$ and 
$\Si_1\cdot L'_1 \cup \Si_2\cdot L'_2$ is an
$\om$-language in  the class {\bf r}-${\bf BCL}(4)_\om$. 

The reverse direction is immediate: if $L \subseteq \Sio$ is a   Borel set accepted by a  $4$-blind counter B\"uchi  automaton $\mathcal{A}$, then it is also accepted by a B\"uchi Turing machine and thus by \cite[Theorem 25]{Fin-mscs06} there exists a real time 
$1$-counter B\"uchi automaton  $\mathcal{B}$ such that $L(\mathcal{A})  \equiv_W L(\mathcal{B})$. 
\ep 

\hs We have only considered in the above Theorem \ref{thewad} the Wadge hierarchy of {\bf  Borel sets}. But we know that there exist also some non-Borel $\om$-languages accepted by 
real time 
$1$-counter B\"uchi automata, and even some ${\bf \Si}_1^1$-complete ones, \cite{Fin03a}.

   By  Lemma 4.7 of \cite{Fin13-JSL} 
the conclusion of the above Lemma \ref{sum-wad} is also true if 
$L$ is assumed to be an analytic but non-Borel set. 

\begin{lem}[\cite{Fin13-JSL}]\label{sum-wad2}
Let $L \subseteq \Sio$ be an analytic but non-Borel set. Then  $L \equiv_W \emptyset + L$. 
\end{lem}

\noi 
      Next  the proof of the above Lemma  \ref{nad} can be adapted to the case of an  analytic but non-Borel set, and we can state the following result. 
  
  \begin{lem}\label{nad2}
Let  $L \subseteq \Sio$ be an analytic but non-Borel set acccepted by a real time 
$1$-counter B\"uchi  automaton   $\mathcal{A}$. 
Then there is an $\om$-language $L'$ accepted by   a real time $4$-blind counter B\"uchi  automaton such that $L \equiv_W L'$.  
\end{lem}
  
 \proo  It is very similar to the proof of the above Lemma  \ref{nad}, using Lemma \ref{sum-wad2} instead of the above Lemma \ref{sum-wad}.
\ep

\hs It was proved in \cite{Fin-mscs06} that the Wadge hierarchy of the class {\bf r}-${\bf CL}(1)_\om$ is equal to the Wadge hierarchy of the class $\Si_1^1$ of 
effective analytic sets. 
Using  Lemma \ref{sum-wad2} instead of  the above Lemma \ref{sum-wad},  the proofs of \cite{Fin-mscs06} can also be adapted  to the case of a non-Borel set to show that 
     for every effective analytic but non-Borel set $L \subseteq \Sio$, where $\Si$ is a finite alphabet, there exists an $\om$-language $L'$ in 
 {\bf r}-${\bf CL}(1)_\om$ such that $L' \equiv_W L$. 

 We can finally summarize our results by the following theorem. 

\begin{The}\label{the-complet}    
The Wadge hierarchy of the class {\bf r}-${\bf BCL}(4)_\om$ is the Wadge hierarchy of the class 
{\bf r}-${\bf CL}(1)_\om$ and also  of the class 
 $\Sigma^1_1$ of effective analytic sets. 
 Moreover for every effective analytic set $L \subseteq \Sio$ there exists an $\om$-language $L'$  in the class  {\bf r}-${\bf BCL}(4)_\om$ such that $L\equiv_W L'$. 
\end{The}

\begin{Rem}
Since the class {\bf r}-${\bf PN}_\om=\bigcup_{k\geq 1}$ {\bf r}-${\bf BCL}(k)_\om$   of $\om$-languages of real-time {\it non-deterministic} Petri nets satisfy the following inclusions 
 {\bf r}-${\bf BCL}(4)_\om$ $\subseteq $ {\bf r}-${\bf PN}_\om$ $\subseteq $ $\Sigma^1_1$, it holds that the  Wadge hierarchy of the class {\bf r}-${\bf PN}_\om$ is also equal to the 
Wadge hierarchy of the class  $\Sigma^1_1$ of effective analytic sets. Moreover the same result  holds for the class ${\bf PN}_\om$ of $\om$-languages of (possibly non-real-time) 
{\it non-deterministic} Petri nets.  
\end{Rem}

 On the other hand, for each non-null countable ordinal $\alpha$ the ${\bf \Si}^0_\alpha$-complete sets 
(respectively, the  ${\bf \Pi}^0_\alpha$-complete  sets) form a single Wadge degree.  Moreover for each non-null recursive ordinal $\alpha < \om_1^{{\rm CK}} $
there are some ${\bf \Si}^0_\alpha$-complete sets and  some ${\bf \Pi}^0_\alpha$-complete  sets in the effective class $\Delta_1^1$. Thus  we can infer the following result
 from the above Theorem \ref{the-complet} and from    the results of 
 \cite{KMS89}. 

\begin{Cor}
For each non-null recursive ordinal $\alpha < \om_1^{{\rm CK}} $ there 
exist some ${\bf \Si}^0_\alpha$-complete and some 
 ${\bf \Pi}^0_\alpha$-complete   $\om$-languages  in the class  {\bf r}-${\bf BCL}(4)_\om$. 
And the supremum 
of the set of Borel ranks of $\om$-languages  in the class  {\bf r}-${\bf BCL}(4)_\om$ is the ordinal $\gamma_2^1$, which  is precisely defined in \cite{KMS89}.  
\end{Cor}

Since it was  proved in \cite{Fin03a} that 
there is a ${\bf \Si}_1^1$-complete set accepted by a  real-time $1$-counter B\"uchi automaton, we also get the following result. 

\begin{Cor}
There exists some ${\bf \Si}_1^1$-complete set in the class {\bf r}-${\bf BCL}(4)_\om$. 
\end{Cor}

Notice that  if we assume the axiom of ${\bf \Si}_1^1$-determinacy, then   any set which is analytic but not Borel is 
${\bf \Si}_1^1$-complete, see \cite{Kechris94}, and thus  there is only one more Wadge degree  (beyond Borel sets) containing  ${\bf \Si}_1^1$-complete sets. 
On the other hand,  if the  axiom of  (effective) ${\Si}_1^1$-determinacy does not hold, then there exist some effective analytic sets which are neither Borel nor  ${\bf \Si}_1^1$-complete. 
Recall that ZFC is  the commonly accepted axiomatic 
framework for Set Theory in which all usual mathematics can be developed.

\begin{Cor}\label{non-complete}
It is consistent with ZFC that there exist some 
$\om$-languages of Petri nets  in the class {\bf r}-${\bf BCL}(4)_\om$     which are neither Borel nor  ${\bf \Si}_1^1$-complete. 
\end{Cor}

    We also prove that it is highly undecidable to determine the topological complexity of a Petri net $\om$-language.  As usual, since there is a finite description of a  real time 
$1$-counter B\"uchi automaton or of a 4-blind-counter  B\"uchi automaton, we can define a G\"odel  numbering of all $1$-counter B\"uchi automata or of all 
 4-blind-counter  B\"uchi automata and then speak about the $1$-counter B\"uchi automaton (or 4-blind-counter  B\"uchi automaton) of index $z$. 
  Recall first the following result,  proved in  \cite{Fin-HI}, where 
  we denote  $\mathcal{A}_z$ the real time 
$1$-counter B\"uchi automaton of index $z$ reading words over a fixed finite alphabet $\Si$ having at least two letters. 
We refer the reader to a textbook like \cite{Odifreddi1,Odifreddi2} 
for more background about the 
analytical hierarchy of subsets of the set $\mathbb{N}$ of natural numbers. 
  
    \begin{The}\label{borel-hard}
\noi Let $\alpha$ be a countable ordinal. Then  
\begin{enumerate}
\ite $ \{  z \in \mathbb{N}  \mid  L(\mathcal{A}_z) \mbox{ is in the Borel class } {\bf \Si}^0_\alpha \}$ is  $\Pi_2^1$-hard. 
\ite  $ \{  z \in \mathbb{N}  \mid  L(\mathcal{A}_z) \mbox{ is in the Borel class } {\bf \Pi}^0_\alpha \}$ is  $\Pi_2^1$-hard. 
\ite  $ \{  z \in \mathbb{N}  \mid  L(\mathcal{A}_z) \mbox{ is a  Borel set } \}$ is  $\Pi_2^1$-hard. 
\end{enumerate}
\end{The}
  
 Using the previous constructions we can now easily show the following result, where $\mathcal{P}_z$ is the real time 4-blind-counter  B\"uchi automaton of indez $z$.

        \begin{The}\label{borel-hard-pn}
\noi Let $\alpha\geq 2$ be a countable ordinal. Then  
\begin{enumerate}
\ite $ \{  z \in \mathbb{N}  \mid  L(\mathcal{P}_z) \mbox{ is in the Borel class } {\bf \Si}^0_\alpha \}$ is  $\Pi_2^1$-hard. 
\ite  $ \{  z \in \mathbb{N}  \mid  L(\mathcal{P}_z) \mbox{ is in the Borel class } {\bf \Pi}^0_\alpha \}$ is  $\Pi_2^1$-hard. 
\ite  $ \{  z \in \mathbb{N}  \mid  L(\mathcal{P}_z) \mbox{ is a  Borel set } \}$ is  $\Pi_2^1$-hard. 
\end{enumerate}
\end{The}

    \proo  It follows  from the fact that one can easily get an injective  recursive function $g: \mathbb{N} \ra \mathbb{N}$ such that 
$\mathcal{P}_{\mathcal{A}_z} =  h( L(\mathcal{A}_z) ) \cup \mathcal{L} = L(\mathcal{P}_{g(z)}) $ and from the following equivalences which hold for each countable ordinal 
$\alpha \geq 2$: 

\begin{enumerate}
\ite  $L(\mathcal{A}_z)$  is in the Borel class  ${\bf \Si}^0_\alpha$  (resp.,  ${\bf \Pi}^0_\alpha$)   $\Longleftrightarrow$ $ L(\mathcal{P}_{g(z)})$  is in the Borel class  ${\bf \Si}^0_\alpha$ 
(resp., ${\bf \Pi}^0_\alpha$).
\ite  $L(\mathcal{A}_z)$  is a  Borel set  $\Longleftrightarrow$ $ L(\mathcal{P}_{g(z)})$  is a  Borel  set. \ep
\end{enumerate}

\section{Concluding remarks}

We have proved  that  the Wadge hierarchy of  Petri nets $\om$-languages, and even of  $\om$-languages in the class {\bf r}-${\bf BCL}(4)_\om$, 
is equal to the Wadge hierarchy of  effective analytic sets, and that it is highly undecidable to determine the topological complexity of a Petri net $\om$-language. 
In some sense our results show that, in contrast with the finite behavior,  the infinite behavior of Petri nets is closer to the infinite behavior of Turing machines than to that of finite automata.

It remains open  for further study to determine the Borel and Wadge hierarchies of   $\om$-languages accepted by automata with less than four blind counters. Since this paper has been written, Micha{\l} Skrzypczak has informed us that he has recently proved in this direction  that there exists a ${\bf \Si}_1^1$-complete $\om$-language  accepted by a 1-blind-counter automaton, \cite{S17}. 
In particular, it then   remains open to determine whether there exist some  $\om$-languages accepted by 1-blind-counter automata which are Borel  of rank greater than $3$,   or which could be  neither Borel nor  ${\bf \Si}_1^1$-complete. 

Finally we mention that, in  an extended version of this paper,  we prove that the determinacy of Wadge games between two players in charge of $\om$-languages of Petri nets is equivalent to the (effective) analytic determinacy,  which is known to be a large cardinal assumption, and thus is not provable in the axiomatic system ZFC. 
Based on the constructions used in the proofs of the above results, we also show that the equivalence and  the inclusion problems for 
$\om$-languages of Petri nets are $\Pi_2^1$-complete, hence highly undecidable.

 \newpage 

\newpage 

 \section*{ANNEXE 1}

\hs \noindent We give in this annexe some proofs which could not be included in the paper, due to lack of space.

\hs \noi {\bf Proof of  Lemma  \ref{lem6}.}   

 \hs \noi 
Assume first that  $y \in L(\mathcal{B})\setminus h(\Sio)$   is of  the following form
$$ y = A0^{n_1}x(1)B0^{n_2}x(2)A \cdots   B0^{n_{2n}}x(2n)A0^{n_{2n+1}}x(2n+1)B \cdots  $$
\noi and that the smallest integer $i$ such that $n_i \neq i$ is an even integer $i_0>1$. 
Consider an infinite accepting run of $\mathcal{B}$ on $y$. It follows from the  proof of the above Lemma \ref{lem5} that after the reading of the initial segment 
$$ A0^{n_1}x(1)B0^{n_2}x(2)A \cdots  
A0^{{i_0-1}}x(i_0-1)B$$
\noi the values of the counters  $\mathcal{C}_3$ and   $\mathcal{C}_4$ are equal to zero, and the values of the counters $\mathcal{C}_1$ and   $\mathcal{C}_2$ satisfy 
$|\mathcal{C}_1| + |\mathcal{C}_2|=i_0-1$. Thus since the two counters must successively decrease during the next $n_ {i_0}-1$  letters $0$, it holds that 
$n_ {i_0}-1 \leq i_0-1$ because otherwise either $\mathcal{C}_1$ or  $\mathcal{C}_2$ would block. Therefore $n_{i_0} < i_0$ since $n_{i_0} \neq  i_0$ by definition of $i_0$. 
The reasoning is very similar in the case of an odd integer $i_0$,  the role  of the counters $\mathcal{C}_1$ and $\mathcal{C}_2$ on one side, and of the counters $\mathcal{C}_3$ and   $\mathcal{C}_4$  on the other side, being simply  interchanged. 
\ep

\hs \noi {\bf Proof of  Corollary  \ref{non-complete}.}   

 \hs \noi 
Recall that ZFC is  the commonly accepted axiomatic 
framework for Set Theory in which all usual mathematics can be developed.  The determinacy of Gale-Stewart games $G(A)$,  where 
$A$ is an (effective) analytic set, denoted {\bf Det}($\Si_1^1$),  is not provable in ZFC; Martin and Harrington have proved that it is  a large cardinal 
assumption equivalent to the existence of a particular real, called the real $0^\sharp$, see \cite[page  637]{Jech}. 
It is also known that the determinacy of (effective) analytic 
Gale-Stewart games is equivalent to the determinacy of (effective) analytic Wadge games, denoted  {\bf W-Det}($\Si_1^1$), see \cite{Louveau-Saint-Raymond}. 

It is known that, if ZFC is consistent, then there is a model of ZFC in which the determinacy of (effective) analytic 
Gale-Stewart games, and thus also the determinacy of (effective) analytic Wadge games, do not hold. It follows from \cite[Theorem 4.3]{harrington} that in such a model of ZFC there exists an effective analytic set which is neither Borel nor 
${\bf \Si}_1^1$-complete. The result now follows from Theorem \ref{the-complet}.

\newpage 

 \section*{ANNEXE 2}

\hs \noindent Some additional results are given here.

We proved in \cite{Fin13-JSL}  that the determinacy of Wadge games between two players in charge of 
$\om$-languages accepted by  real time $1$-counter B\"uchi automata,  denoted    {\bf W-Det}({\bf r}-${\bf CL}(1)_\om$),   is  equivalent to the  (effective) analytic  Wadge determinacy. 

We can now state the following result, proved within the axiomatic system ZFC.

\begin{The}
The determinacy of Wadge games between two players in charge of $\om$-languages in the class  {\bf r}-${\bf BCL}(4)_\om$ is equivalent to the effective analytic (Wadge) determinacy, and thus  is not provable in  the axiomatic system ZFC. 
\end{The}

\proo  It was proved in \cite{Fin13-JSL} that  the following equivalence holds: {\bf W-Det}({\bf r}-${\bf CL}(1)_\om$) $\Longleftrightarrow$  {\bf W-Det}($\Si_1^1$).  
The implication {\bf W-Det}($\Si_1^1$)$\Longrightarrow$ {\bf W-Det}({\bf r}-${\bf BCL}(4)_\om$) is obvious since the class ${\bf BCL}(4)_\om$  is included into the class $\Si_1^1$. 
To prove the reverse implication, we assume that {\bf W-Det}({\bf r}-${\bf BCL}(4)_\om$) holds and we show that every Wadge game $W(L(\mathcal{A}),L(\mathcal{B}))$  between two players in charge of 
$\om$-languages of the class {\bf r}-${\bf CL}(1)_\om$ is determined (we assume without loss of generality that the two real time 1-counter  B\"uchi automata $\mathcal{A}$ and $\mathcal{B}$ read words over the same alphabet $\Si$). 

It is sufficient to consider the cases where at least one of two $\om$-languages $L(\mathcal{A})$ and $L(\mathcal{B})$ is non-Borel, since the Borel Wadge determinacy is provable in 
ZFC.  On the other hand, we have seen how we can construct some real time 4-blind-counter  B\"uchi automata $\mathcal{P}_{\mathcal{A}}$ and $\mathcal{P}_{\mathcal{B}}$
such that $L(\mathcal{P}_{\mathcal{A}})=h(L(\mathcal{A})) \cup \mathcal{L}$  and $L(\mathcal{P}_{\mathcal{B}})=h(L(\mathcal{B})) \cup \mathcal{L}$. 

We can firstly  consider the case where  $L(\mathcal{A})$ is Borel of Wadge degree smaller than $\om$, and $L(\mathcal{B})$  is non-Borel. In that case $L(\mathcal{A})$ is in particular a 
${\bf \Pi}^0_2$-set.  Recall now that we can infer from Hurewicz's Theorem, see \cite[page 160]{Kechris94}, 
that an analytic subset of $\Si^\om$  is either ${\bf \Pi}^0_2$-hard   or  a ${\bf \Si}^0_2$-set. Thus $L(\mathcal{B})$  is ${\bf \Pi}^0_2$-hard and Player 2 has a winning strategy 
in the game  $W(L(\mathcal{A}),L(\mathcal{B}))$. 

Secondly we consider the case where  $L(\mathcal{A})$ and $L(\mathcal{B})$ are either non-Borel or Borel of Wadge degree greater  than $\om$. 
By hypothesis we know that the Wadge game $W(L(\mathcal{P}_{\mathcal{A}}), L(\mathcal{P}_{\mathcal{B}}))$ is determined, and that one of the players has a winning strategy. 
Using the above constructions and reasonings we used in the proofs of Lemmas \ref{lem8} and \ref{nad}, 
we can easily show that  the same player has a winning strategy in the Wadge game $W(L(\mathcal{A}),L(\mathcal{B}))$. 

 We now consider the two following cases: 
\nl {\bf First case.} Player 2 has a w.s. in the game $W(L(\mathcal{P}_{\mathcal{A}}), L(\mathcal{P}_{\mathcal{B}}))$. If $L(\mathcal{B})$ is Borel then $L(\mathcal{P}_{\mathcal{B}})$ is easily seen to be Borel 
 and then $L(\mathcal{P}_{\mathcal{A}})$ is also Borel because $L(\mathcal{P}_{\mathcal{A}}) \leq_W    L(\mathcal{P}_{\mathcal{B}})$.  Thus  $L(\mathcal{A})$ is also Borel and  the game 
$W(L(\mathcal{A}), L(\mathcal{B}))$ is determined. Assume now that $L(\mathcal{B})$ is not Borel. Consider the Wadge game $W(L(\mathcal{A}), \emptyset + (\emptyset + L(\mathcal{B})))$. We claim that Player 2 has a w.s. in that 
game which is easily deduced from a w.s. of Player 2 in the Wadge game  $W(L(\mathcal{P}_{\mathcal{A}}), L(\mathcal{P}_{\mathcal{B}}))$ $=W(h(L(\mathcal{A})) \cup \mathcal{L}, h(L(\mathcal{B})) \cup \mathcal{L})$. Consider a play in this latter game where  Player 
1 remains in the closed set $h(\Si^\om)$:  she writes a beginning of a word in the form 
$$A0x(1)B0^{2}x(2)A \cdots  
B0^{2n}x(2n)A \cdots  $$
\noi Then player 2 writes a beginning of a word in the form 
$$A0x'(1)B0^{2}x'(2)A \cdots  
B0^{2p}x'(2p)A \cdots $$
\noi where $p\leq n$. 
Then the strategy for Player 2 in $W(L(\mathcal{A}), \emptyset + (\emptyset + L(\mathcal{B})))$ consists to write $x'(1).x'(2) \ldots 
x'(2p).$ when Player 1 writes  $x(1).x(2) \ldots x(2n)$. (Notice that Player 2 is allowed to skip, provided he really writes an $\om$-word in $\om$ steps). 
If the strategy for Player 2 in $W(L(\mathcal{P}_{\mathcal{A}}), L(\mathcal{P}_{\mathcal{B}}))$ was at some step to go out of
the  closed set $h(\Si^\om)$ then this means that   the word he has now 
written is not a prefix of any $\om$-word of  $h(\Sio)$,  and  Player 2 ``enters'' in the open set $h(\Sio)^- = \mathcal{L}\cup \mathcal{L}'$ and will stay in this set. Two subcases may now appear. 

{\bf Subcase A.} When  Player 2  in the game $W(L(\mathcal{P}_{\mathcal{A}}), L(\mathcal{P}_{\mathcal{B}}))$  ``enters'' in the open set $h(\Sio)^- = \mathcal{L}\cup \mathcal{L}'$, he actually enters in the open set $ \mathcal{L}$. Then the final word written by Player 2 will surely be inside $L(\mathcal{P}_{\mathcal{B}})$. Player 2 in the Wadge game $W(L(\mathcal{A}), \emptyset + (\emptyset + L(\mathcal{B})))$ can now  write a letter of $\Delta -\Si$ in such a way that he is now like a player in charge of the wholeset  and he can 
now writes an $\om$-word $u$ so that his final $\om$-word will be inside $\emptyset +  (\emptyset + L(\mathcal{B}))$. Thus Player 2 wins this play too. 

{\bf  Subcase B.}   When  Player 2  in the game $W(L(\mathcal{P}_{\mathcal{A}}), L(\mathcal{P}_{\mathcal{B}}))$  ``enters'' in the open set $h(\Sio)^- = \mathcal{L}\cup \mathcal{L}'$, he does not enter in the open set $ \mathcal{L}$. 
Then  Player 2, in the Wadge game $W(L(\mathcal{A}), \emptyset + (\emptyset + L(\mathcal{B})))$,  being first like a player in charge of the set $(\emptyset +  L(\mathcal{B}))$,  can write a  letter of $\Delta -\Si$ in such a way that he is now like a player in charge of the emptyset  and he can 
now continue, writing an $\om$-word $u$. If Player 2  in the game $W(L(\mathcal{P}_{\mathcal{A}}), L(\mathcal{P}_{\mathcal{B}}))$  never enters  in the open set $\mathcal{L}$ then the final word written by Player 2 will be in $\mathcal{L}'$ and thus surely outside $L(\mathcal{P}_{\mathcal{B}})$, and the final word written by Player 2 will be outside the emptyset. So in that case Player 2 wins this play too in the Wadge game $W(L(\mathcal{A}), \emptyset + (\emptyset + L(\mathcal{B})))$. 
If at some step of the play, in the game $W(L(\mathcal{P}_{\mathcal{A}}), L(\mathcal{P}_{\mathcal{B}}))$,  Player  2 enters  in the open set $\mathcal{L}$ then his final $\om$-word will be surely in $L(\mathcal{P}_{\mathcal{B}}))$. In that case Player 2, in charge of the set 
$\emptyset + (\emptyset + L(\mathcal{B})))$   in the Wadge game $W(L(\mathcal{A}), \emptyset + (\emptyset + L(\mathcal{B})))$, can again write an extra letter and choose to be  in charge of the wholeset and he can 
now write an $\om$-word $v$ so that his final $\om$-word will be inside $\emptyset + (\emptyset +  L(\mathcal{B})))$. Thus Player 2 wins this play too.

  So we have proved that Player 2 has a w.s. in  the Wadge game $W(L(\mathcal{A}), \emptyset + (\emptyset + L(\mathcal{B})))$  or equivalently that 
$L(\mathcal{A}) \leq_W   \emptyset + (\emptyset + L(\mathcal{B}))$. But by  Lemma \ref{} we know that  $L(\mathcal{B})  \equiv_W \emptyset + (\emptyset + L(\mathcal{B}))$  and thus 
$L(\mathcal{A}) \leq_W L(\mathcal{B})$ which means that Player 2 has a  w.s. in  the Wadge game $W(L(\mathcal{A}), L(\mathcal{B}))$.

\hs {\bf Second case.} Player 1 has a w.s. in the game $W(L(\mathcal{P}_{\mathcal{A}}), L(\mathcal{P}_{\mathcal{B}}))$. Notice that this implies that 
$ L(\mathcal{P}_{\mathcal{B}}) \leq_W L(\mathcal{P}_{\mathcal{A}})^-$. Thus if $L(\mathcal{A})$ is Borel then $L(\mathcal{P}_{\mathcal{A}})$ is Borel, $L(\mathcal{P}_{\mathcal{A}})^-$ is also Borel, 
and  $ L(\mathcal{P}_{\mathcal{B}})$ is Borel as the inverse image of a Borel set by a continuous function, and $ L(\mathcal{B})$ is also Borel, so the Wadge game 
$W( L(\mathcal{A}),  L(\mathcal{B}))$ is determined. We now  assume that $L(\mathcal{A})$ is not Borel and we consider the Wadge game $W(L(\mathcal{A})),  L(\mathcal{B}))$. 
Player 1 has a w.s. in this game which is easily constructed from a w.s. of the same player in the game  $W(L(\mathcal{P}_{\mathcal{A}}), L(\mathcal{P}_{\mathcal{B}}))$ as follows. 
For this consider a play in this latter game where Player 2 does not go out of the closed set $h(\Sio)$. 

He writes a beginning of a word in the form 
$$A0x(1)B0^{2}x(2)A \cdots  
B0^{n}x(n)A \cdots  $$
\noi Then player 1 writes a beginning of a word in the form 
$$A0x'(1)B0^{2}x'(2)A \cdots  
B0^{p}x'(p)A \cdots $$
where $n \leq p$ (notice that here without loss of generality  the notation implies that $n$ and $p$ are even, since the segments $B0^{n}x(n)A$ and $B0^{p}x'(p)A$ begin
 with a letter $B$      but this is not essential in the proof). 
Then the strategy for Player 1 in $W(L(\mathcal{A})),  L(\mathcal{B})$ consists to write $x'(1).x'(2) \ldots 
x'(p).$ when Player 2 writes  $x(1).x(2) \ldots x(n)$.   After $\om$ steps, the $\om$-word written by Player 1 is in $L(\mathcal{A})$ iff the $\om$-word written by Player 2 is not in the set 
 $L(\mathcal{B})$, and thus Player 1 wins the play. 

If the strategy for Player 1 in $W(L(\mathcal{P}_{\mathcal{A}}), L(\mathcal{P}_{\mathcal{B}}))$ was at some step to go out of
the  closed set $h(\Si^\om)$ then this means that  she ``enters'' in the open set $h(\Sio)^- = \mathcal{L}\cup \mathcal{L}'$ and will stay in this set. Two subcases may now appear.

{\bf Subcase A.} When  Player 1  in the game $W(L(\mathcal{P}_{\mathcal{A}}), L(\mathcal{P}_{\mathcal{B}}))$  ``enters'' in the open set $h(\Sio)^- = \mathcal{L}\cup \mathcal{L}'$, she actually enters in the open set $ \mathcal{L}$. Then the final word written by Player 1 will surely be inside $L(\mathcal{P}_{\mathcal{A}})$. But she wins the play since she follows a winning strategy and this leads to a contradiction. Indeed if Player 2 decided to also enter in  in the open set $ \mathcal{L}$ then Player 2 would win the play. Thus this case is actually not possible. 

{\bf  Subcase B.}   When  Player 1  in the game $W(L(\mathcal{P}_{\mathcal{A}}), L(\mathcal{P}_{\mathcal{B}}))$  ``enters'' in the open set $h(\Sio)^- = \mathcal{L}\cup \mathcal{L}'$, she does not enter in the open set $ \mathcal{L}$. But Player 2 would be able to do the same and enter in $h(\Sio)^- = \mathcal{L}\cup \mathcal{L}'$ but not (for the moment) in  the open set $ \mathcal{L}$. And if at some step of the play, Player 1 would enter in the open set $ \mathcal{L}$ then Player 2 could do the same, and thus Player 2 would win the play. 
Again this is not possible since Player 1 wins the play since she follows a winning strategy.

 Finally both subcases A and B cannot occur and this shows that Player 1 has a w.s. in the Wadge game $L(\mathcal{A}) \leq_W L(\mathcal{B})$. 
\ep

\hs We now add the following undecidability results. We first recall  the following result,  proved in  \cite{Fin-HI}. 

\begin{The}[\cite{Fin-HI}]\label{the-ind}
The  equivalence and the inclusion problems for $\om$-languages accepted by 
real time $1$-counter B\"uchi automata are $\Pi_2^1$-complete, i.e.:
\begin{enumerate}
\ite $\{ (z,z')\in \mathbb{N} \mid L(\mathcal{A}_z) = L(\mathcal{A}_{z'}) \}$ is $\Pi_2^1$-complete
\ite $\{ (z,z')\in \mathbb{N} \mid L(\mathcal{A}_z) \subseteq L(\mathcal{A}_{z'}) \}$ is $\Pi_2^1$-complete
\end{enumerate}
\end{The}

Using the previous constructions we can now easily show the following result, where $\mathcal{P}_z$ is the real time 4-blind-counter  B\"uchi automaton of indez $z$. 

\begin{The}\label{}
The equivalence and the inclusion problems for $\om$-languages of Petri nets, or even for   $\om$-languages in the class   {\bf r}-${\bf BCL}(4)_\om$,    are $\Pi_2^1$-complete. 
\begin{enumerate}
\ite $\{ (z,z')\in \mathbb{N} \mid L(\mathcal{P}_z) = L(\mathcal{P}_{z'}) \}$ is $\Pi_2^1$-complete
\ite $\{ (z,z')\in \mathbb{N} \mid L(\mathcal{P}_z) \subseteq L(\mathcal{P}_{z'}) \}$ is $\Pi_2^1$-complete
\end{enumerate}

\end{The}

\proo Firstly, it is easy to see that each of these decision problems is in the class $\Pi_2^1$, since the equivalence and the inclusion problems for $\om$-languages of Turing machines are already in the class $\Pi_2^1$, see \cite{cc,Fin-HI}. 
The completeness part follows from the above Theorem \ref{the-ind},  from the fact that there exists an injective  recursive function $g: \mathbb{N} \ra \mathbb{N}$ such that 
$\mathcal{P}_{\mathcal{A}_z}=\mathcal{P}_{g(z)}$, and then from the following equivalences:

\begin{enumerate}
\ite  $L(\mathcal{A}_z) = L(\mathcal{A}_{z'})   \Longleftrightarrow  L(\mathcal{P}_{g(z)}) = L(\mathcal{P}_{g(z')})$ 
\ite  $L(\mathcal{A}_z) \subseteq L(\mathcal{A}_{z'}) 
\Longleftrightarrow L(\mathcal{P}_{g(z)}) \subseteq L(\mathcal{P}_{g(z')})$
\end{enumerate}

which clearly imply that the equivalence (respectively, inclusion) problem for $\om$-languages of real-time 1-counter automata is Turing reducible to the 
equivalence (respectively, inclusion) problem for $\om$-languages of  real time 4-blind-counter  B\"uchi automata.
  \ep

\end{document}